\documentstyle[prd,aps]{revtex}

\input epsf.sty

\preprint{UCTP-106-98}  

\title{Derivative Expansion of the Effective Action for QED
in 2+1 and 3+1 dimensions}

\author{V.P.~Gusynin}

\address{Bogolyubov Institute for Theoretical Physics,
252143, Kiev, Ukraine\\
and \\
Institute for Theoretical Physics, University of Bern, 
Sidlerstrasse 5, CH-3012 Bern, Switzerland}

\author{I.A.~Shovkovy\thanks{On leave of absence from 
Bogolyubov Institute for Theoretical Physics, 252143, 
Kiev, Ukraine.}}

\address{Physics Department, University of Cincinnati,
Cincinnati, Ohio 45221-0011}

\draft

\begin{document}

\maketitle

\begin{abstract}
The derivative expansion of the one-loop effective action in
QED$_3$ and QED$_4$ is considered. The first term in such an
expansion is the effective action for a constant electromagnetic
field. An explicit expression for the next term containing two
derivatives of the field strength $F_{\mu\nu}$, but exact in the
magnitude of the field strength, is obtained. The general results 
for both fermion and scalar electrodynamics are presented. The 
cases of pure electric and pure magnetic external fields are 
considered in detail. The Feynman technique for the perturbative 
expansion of the one-loop effective action in the number of 
derivatives is developed.
\end{abstract}


\section{Introduction}

Quantum electrodynamics is known to be the best studied example of 
quantum field theory. Mainly, this is due to the weakness of the 
fine structure (coupling) constant, $\alpha\approx 1/137$, what
allows to perform many perturbative calculations as power series
in $\alpha$ with an incredibly high accuracy. Despite the
smallness of  $\alpha$, even in the realm of quantum
electrodynamics, there are some  questions that theory has not
answered yet. In this paper, in particular,  we address the
problem of derivation of the low-energy effective action  which at
present is solved only partially for QED. 

The low-energy effective action in quantum electrodynamics 
describes the dynamics of the electromagnetic field, assuming  
that the production of the on shell fermions is absent or
negligible. Apparently, such a description is self-consistent only
if the fermions are massive and the characteristic photon energies
are sufficiently  small. The mentioned two conditions, as is
clear, are necessary to  suppress the process of the
particle-antiparticle pair creation  (on-shell).
 
Intuitively, the low-energy effective theory is obtained from
quantum  electrodymanics by ``integrating out" the fermion field.
After doing so,  one arrives at a nonlinear theory that involves
only the electromagnetic  field degrees of freedom. In terms of
the S-matrix language, one considers just those processes in QED
which contain only photons among the asymptotic  scattering
states. The fermions, on the other hand, appear only through the
internal loops by producing all kinds of photon vertices.

The problem of deriving the effective action is an old one. Its
roots  go back to the well known papers of Heisenberg and Euler
\cite{Heis},  and Weisskopf \cite{Wie}. There, for the first time,
the effective action  in QED (for the case of a constant
electromagnetic field) was derived.  From the viewpoint of
application, the derived effective action contains,  for example,
the information on the photon-photon scattering at the tree level.
It was this scattering process, in fact, that motivated 
consideration of the problem in Ref.~\cite{Heis,Wie}, in the first
place. Later, some further progress was achieved by Schwinger
\cite{Sch} who,  by using the proper time technique, rederived the
result of  Refs.~\cite{Heis,Wie} and, in addition, gave a nice
interpretation to  the imaginary part of the effective action in
the case of a constant  electric field. 

Obviously, the next most natural step in deriving the low-energy
effective  action in QED would be to take into account the effect
of small deviations  from the constant configuration of the field.
In other words, the problem  is to obtain the effective action as
an expansion in powers of derivatives  of the field strength. It
turns out, however, that the latter is very difficult to
accomplish (see \cite{Hauk,Martin} for some early attempts in this
direction) unlesss the weak field approximation is used. In this
connection it is appropriate to mention that, in the weak field
limit, the expansion is known up to four derivatives with respect
to the field strength \cite{Dicus}. Our approach, on the other
hand, does not involve any assumptions about the weakness of the
background field.

A real progress in solving the problem started with the result of 
Ref.~\cite{Lee} where an elaborated method, which, in principle,
leads to a general result for the derivative expansion in QED, was
presented. Because of the complicated character of the method,
however, the explicit expression applicable to the most general
case of the electromagnetic field background was not presented
there. Recently, the derivative expansion of the effective  action
was obtained in the case of $(2+1)$-dimensional QED \cite{DHok}.
This latter is a quite general result, containing all the terms
quadratic in derivatives of the field strength with respect to the
space-time coordinates. Finally, in our previous paper \cite{GS},
we obtained a similar result for the effective action but in
$(3+1)$-dimensional QED. As in $(2+1)$-dimensional case, it was
given in a covariant form valid for the most general constant
component of the electromagnetic field background what, as we
will see later, is much more complicated  problem than that in 2+1
dimensions. 

For completeness, we mention that some related interesting 
results were obtained in Ref.~\cite{DH} for QED and in
Refs.~\cite{McK,Gargett} for non-Abelian gauge theories. 

In this paper we extend our method, which was originally presented
for the case of $(3+1)$-dimensional QED \cite{GS}, to QED in $2+1$
dimensions. In particular, we obtain the derivative expansion of
the effective action which includes up to two space-time
derivatives of the electromagnetic field and, further, we
formulate the Feynman rules for the perturbative expansion of the
one-loop effective action in the number of derivatives. We also
derive the explicit expressions for the derivative  corrections to
the imaginary part of the effective action in an external electric
field. And finally, as a byproduct, we resolve the controversy
posed in \cite{DitGi} where a result different from that of
\cite{DHok} was presented.

The paper is organized as follows. In Sec.~\ref{GenFor} we
outline the general method developed in our previous paper
\cite{GS}. Sec.~\ref{Matr} is devoted to solving some technical
problems in dealing with functions of the matrix argument $F_{\mu
\nu }$. Then, in Secs.~\ref{GenerF} and \ref{GenerS}, we present
the main results of our paper, namely, the derivative expansions
for spinor and scalar QED, respectively. In
Secs.~\ref{QED(2+1)F}-\ref{QED(3+1)F} and
Sec.~\ref{QED(2+1)S}-\ref{QED(3+1)S} we calculate the derivative 
expansions for two particular cases of the external
electromagnetic field, the purely magnetic and purely electric
backgrounds, in both $2+1$ and $3+1$ dimensions. Finally, in
Sec.~\ref{Feynman}, we develop the Feynman diagram technique for
generating the perturbative expansion in the number of
derivatives. Four appendices contain different formulas used
throughout the main text.

\section{Derivative Expansion of the One-Loop 
     Effective Action in QED}
\label{GenFor}

Let us start from the general formalism which was originally
developed in \cite{GS} for $(3+1)$-dimensional quantum
electrodynamics. While doing so, we will notice that, to a great
extent, the method does not depend on the dimension of the
space-time. We will pay special attention to all those places
where it does depend.

In this paper we restrict ourselves to the one-loop effective
action. This is the same approximation which was used by Schwinger
\cite{Sch} in the case of a constant external electromagnetic 
field. 

As is known, the one-loop effective action in QED reduces to 
computing the fermion determinant
\begin{eqnarray}
W^{(1)}(A)&\equiv & \int d^{n}x{\cal L}^{(1)}
=-i\ln {\rm Det}(i\hat{{\cal D}}-m)=
-\frac{i}{2}\ln{\rm Det}\left({\cal D}^2_{\mu}
+\frac{e}{2}\sigma_{\mu\nu}F^{\mu\nu}+m^2\right)=\nonumber\\
&=&-\frac{i}{2}\int d^{n}x\langle x|tr\ln\left({\cal D}^2_{\mu}+
\frac{e}{2}\sigma_{\mu\nu}F^{\mu\nu}+m^2\right) |x\rangle.
\label{eq:ActGen2}
\end{eqnarray}
Here $\hat{{\cal D}}=\gamma^{\mu}{\cal D}_{\mu}$ and the covariant
derivative is ${\cal D}_{\mu}= \partial_{\mu}+ieA_{\mu}$. By
definition, $\sigma_{\mu\nu}=i[\gamma_{\mu},\gamma_{\nu}]/2$ and 
$tr$ refers to the spinor indices of the Dirac matrices
$\gamma_{\mu}$. States $|x\rangle$ are the eigenstates of a
self-conjugate coordinate operator $x_{\mu}$. Throughout the paper
we use the Minkowski metric, i.e., $\eta_{\mu\nu}=(1,-1,-1)$ or
$\eta_{\mu\nu}=(1,-1,-1,-1)$, depending on the actual space-time
dimension. And in both $2+1$ and $3+1$ dimensions, we work with 
the $4\times 4$ representation of the Dirac $\gamma$-matrices. 

For calculating the effective action in Eq.~(\ref{eq:ActGen2}), we 
employ a version of the so-called worldline (or string-inspired) 
formalism developed in \cite{Bern1,Bern2,Stra}. Such an approach
to an ordinary field theory, based on the path integral over 
one-dimensional world lines, was extended to the evaluation of 
Feynman diagrams for Green functions in higher loop orders
\cite{McKeon,schmidt,kors}. It has demonstrated its power 
reproducing known theoretical results in QED while allowing one 
to invoke new technique to study the theory's behavior in strong 
coupling regime \cite{andreas}. For some recent applications 
of the worldline formalism as well as for an extensive 
list of references see \cite{Holt,Reut}. Note, however, that our
method differs from the one commonly used in the literature by a
choice of the worldline propagators, and is closer in spirit to the
method used in \cite{McKeon,McK&Sher}. 

With use of the formal identity $\ln (H+m^2)=-\int_0^{\infty}
\exp[-i\tau (H+m^2)]d\tau /\tau$ for introducing the proper-time 
coordinate $\tau$, the effective Lagrangian can be represented 
through the diagonal matrix elements of the operator 
$U(\tau)=\exp(-i\tau H)$,
\begin{equation}
{\cal L}^{(1)}(A)=\frac{i}{2} \int \limits^{\infty }_{0} 
\frac{d\tau}{\tau} e^{-im^2\tau} tr \langle x| \exp(-i\tau H) 
|x\rangle,
\label{eq:ActGen3}
\end{equation}
where the second order differential operator $H$ is given by
\begin{equation}
H= - \Pi_{\mu} \Pi^{\mu} + \frac{e}{2} \sigma_{\mu\nu} 
F^{\mu\nu}(x) , \qquad \Pi_{\mu} = -i {\cal D}_{\mu}.
\label{eq:operH}
\end{equation}
The matrix elements $\langle x|\exp(-i\tau H)|x\rangle$ entering 
the right hand side of Eq.~(\ref{eq:ActGen3}) may be interpreted 
as the matrix elements of the evolution operator of a spinning
particle with $\tau$ and $H$ being the proper time and the
Hamiltonian of the particle. The corresponding canonical momenta
are $P_{\mu}$'s which obey the commutation relations $[x_{\mu},
P^{\nu}] = i\delta^{\nu}_{\mu}$ and are defined by $\langle x|
P_{\mu} |y\rangle =-i\partial_{\mu} \delta(x-y)$ in coordinate
representation. Following the standard approach \cite{Frad}, we
represent the transition amplitude $\langle z| U(\tau) |y\rangle$
between points $x(0)=y$ and $x(\tau)=z$ in terms of a path integral
over the real and Grassmann coordinates, $x_{\mu}(t)$ and
$\psi_{\mu}(t)$, as 
\begin{equation}
tr \langle z|U(\tau) |y\rangle = N^{-1}\int {\cal D} [x(t) , 
\psi(t) ] \exp \left \{ i\int \limits^{\tau}_{0} dt \left[ L_{bos} 
(x(t)) + L_{fer} (\psi(t) , x(t) ) \right] \right\},
\label{eq:evol}
\end{equation}
where $N$ is a normalization factor, and
\begin{equation}
L_{bos}(x)= -\frac{1}{4}\frac{dx_{\nu }}{dt}
\frac{dx^{\nu }}{dt}
-eA_{\nu }(x) \frac{dx^{\nu }}{dt},
\label{eq:L_bos}
\end{equation}
\begin{equation}
L_{fer}(\psi ,x)= \frac{i}{2} \psi _{\nu }\frac{d\psi ^{\nu }}{dt}-
ie \psi ^{\nu }\psi ^{\lambda } F_{\nu \lambda }(x).
\label{eq:L_fer}
\end{equation}
The integration in Eq.~(\ref{eq:evol}) goes over trajectories 
$x^{\mu}(t)$ and $\psi ^{\mu }(t)$ parameterized by $t\in
[0,\tau]$. The definition of the integration measure assumes the
following boundary conditions
\begin{equation}
x(0)=y,\qquad x(\tau)=z,\qquad \psi (0)=-\psi (\tau).
\end{equation}
We choose a special gauge condition for the vector potential $A_{\mu}(x)$,
namely the Fock-Schwinger gauge \cite{Fock}
\begin{equation}
(x^{\nu }-y^{\nu })A_{\nu }(x)=0,
\end{equation}
which leads to the series
\begin{eqnarray}
A_{\nu }(x) &=& \frac{1}{2} (x^{\lambda } -y^{\lambda }) 
F_{\lambda \nu }(y) + \frac{1}{3} (x^{\lambda }-y^{\lambda })
(x^{\sigma }-y^{\sigma }) \partial _{\sigma } F_{\lambda \nu }(y)
\nonumber\\ &+& 
\frac{1}{8} (x^{\lambda }-y^{\lambda }) (x^{\sigma }-y^{\sigma }) 
(x^{\mu }-y^{\mu }) \partial _{\sigma }\partial _{\mu } 
F_{\lambda \nu }(y) + \ldots \nonumber\\ &=&
\sum^{\infty}_{n=0} \frac{(x^{\lambda }-y^{\lambda }) 
(x^{\nu _{1}}- y^{\nu _{1}}) \ldots (x^{\nu_{n}} - y^{\nu _{n}})}
{n!(n+2)} \partial _{\nu _{1}} \partial _{\nu_2} \ldots 
\partial _{\nu _{n}} F_{\lambda \nu }(y).
\label{eq:Anu}
\end{eqnarray}
This choice of the gauge for the vector potential turns out to be
very convenient for developing a perturbative theory in the
number of the derivatives of the electromagnetic field with respect
to the space-time coordinates. 

Carrying out the change of the variable $x(t)$ for $x'(t)=x(t)-y$
in the path integral in Eq.~(\ref{eq:evol}) (henceforth we omit the
prime) and substituting Eq.~(\ref{eq:Anu}) into Eq.~(\ref{eq:evol}),
we obtain 
\begin{eqnarray} 
tr\langle z|U(\tau)|y\rangle&=&N^{-1}\int D[x(t),\psi (t)] 
\exp\Bigg[i\int\limits^\tau_0 dt \Bigg(-\frac{1}{4}
\frac{d x_\nu}{dt}\frac{d x^\nu}{dt} -\frac{e}{2}x^{\lambda }
F_{\lambda \nu }(y)
\frac{d x^\nu}{dt}+L_{bos}^{int}(x)\Bigg)\Bigg]
\nonumber\\ &\times&\exp\Bigg[i\int\limits^{\tau}_{0}dt
\Bigg(\frac{i}{2}\psi_{\nu} \frac{d\psi^\nu}{dt}-ie\psi^{\nu }
\psi^{\lambda }F_{\nu \lambda }(y) +L_{fer}^{int}(x,\psi 
)\Bigg)\Bigg]. \label{eq:trU1}
\end{eqnarray}
The new boundary conditions for $x(t)$ are $x(0)=0$ and
$x(\tau)=z-y$. Notice, that $F_{\mu \nu }$ in Eq.~(\ref{eq:trU1}) 
does not depend on $x(t)$. As follows from Eqs.~(\ref{eq:L_bos}),
(\ref{eq:L_fer}) and (\ref{eq:Anu}), the expressions for the
interacting terms, $L_{bos}^{int}(x)$ and $L_{fer}^{int}(x,\psi)$,
containing derivatives of $F_{\mu\nu}$ with respect to coordinates,
take the form
\begin{eqnarray}
L_{bos}^{int}(x)&=& \sum^{\infty}_{n=1}\frac{eF_{\nu _{0} \nu _{1}, 
\nu _{2} \ldots \nu _{n+1}}}{n! (n + 2)}\frac{d x^{\nu_0}}{dt}
x^{\nu _{1}} (t)\ldots x^{\nu_{n+1}}(t) \nonumber\\ &=&\frac{e}{3}
F_{\nu \lambda ,\sigma }\frac{d x^\nu}{dt} x^{\lambda } x^{\sigma }
+ \frac{e}{8}F_{\nu \lambda ,\sigma \kappa } \frac{d x^\nu}{dt} 
x^{\lambda } x^{\sigma } x^{\kappa } + \ldots,
\label{eq:L2}\\
L_{fer}^{int}(x,\psi )&=&-\sum^{\infty}_{n=1}\frac{i}{n!}
eF_{\lambda \mu , \nu _{1} \ldots \nu _{n}} \psi ^{\lambda }(t)
\psi ^{\mu }(t) x^{\nu _{1}}(t)\ldots x^{\nu_{n}}(t)
\nonumber\\ &=&-ieF_{\nu 
\lambda ,\sigma } \psi ^{\nu }\psi ^{\lambda }x^{\sigma }- 
\frac{ie}{2}F_{\nu \lambda ,\sigma \kappa } \psi ^{\nu }\psi 
^{\lambda } x^{\sigma }x^{\kappa }+ \ldots.
\label{eq:L3}
\end{eqnarray}
Here we use the conventional notation for the partial derivatives
\begin{eqnarray}
F_{\lambda\mu,\nu _{1}\nu_{2}\ldots\nu _{n}}(x)
=\partial_{\nu_{1}} \partial_{\nu_2}\ldots 
\partial_{\nu_{n}} F_{\lambda\mu}(x).
\end{eqnarray}
Now we see that the problem of obtaining the derivative expansion
reduces to the evaluation of the path integral in
Eq.~(\ref{eq:trU1}) in the framework of the perturbative theory with
an infinite number of interacting terms given in Eqs.~(\ref{eq:L2})
and (\ref{eq:L3}). Fortunately, for computing the effective action
that includes only a finite number of the derivatives, it is
sufficient to consider only a finite number of the interacting
terms. Later, we shall restrict ourselves to obtaining only the
two-derivative terms in the action. So far, we continue developing
the scheme for the most general case.

As usual, introducing real and Grassmann external sources, the 
matrix elements of the evolution operator can be represented as 
follows
\begin{eqnarray}
tr\langle z|U(\tau)|y\rangle&=&
\exp \left\{i \int\limits^{\tau}_{0} dt 
\left[L_{bos}^{int}\left(\frac{1}{i}
\frac{\delta}{\delta \eta (t)}\right) 
+L_{fer}^{int}\left(\frac{1}{i}
\frac{\delta}{\delta \eta (t)},
-\frac{\delta}{\delta \xi (t)} \right)\right]\right\}
\nonumber\\ &\times&
Z_{\tau}[\eta ,\xi ](z;y) \Bigg|_{\eta =0,\xi =0},
\label{eq:trU2}
\end{eqnarray}
where the generating functional is just the Gaussian path 
integral,
\begin{eqnarray}
Z_{\tau}[\eta ,\xi ](z;y)&=&N^{-1}\int D[x(t),\psi (t)]\exp\Bigg[ 
\frac{i}{2} \int \limits^\tau_0 dt\Bigg(-\frac{1}{2} \frac{d 
x_\nu}{dt} \frac{d x^\nu}{dt} - ex^{\lambda }F_{\lambda \nu }(y)
\frac{d x^\nu}{dt}+2\eta_{\nu}x^{\nu}\Bigg)\Bigg]
\nonumber\\ &\times& 
\exp \Bigg[ -\frac{1}{2} \int \limits^{\tau}_{0} dt \Bigg 
(\psi_{\nu} \frac{d\psi^\nu}{dt} -2e \psi^{\nu } \psi^{\lambda }
F_{\nu \lambda }(y) +2\xi_{\nu}\psi^{\nu}\Bigg)\Bigg].
\label{eq:Gfunct}
\end{eqnarray}
The calculation of this generating functional reduces to
obtaining the ``classical" trajectories for $x_{\nu}(t)$ 
and $\psi_{\nu}(t)$, satisfying the appropriate 
boundary conditions, and to computing the determinants
of the one-dimensional differential operators,
\begin{eqnarray}
O_1= \frac{\eta_{\mu\nu}}{2} \frac{d^2}{dt^2} - eF_{\mu\nu} 
\frac{d}{dt}, \qquad \mbox{and}\qquad 
O_2= i\eta_{\mu\nu} \frac{d}{dt} - 2ie F_{\mu\nu},
\label{eq:operas}
\end{eqnarray}
defined on the interval $[0,\tau]$ with the periodic and 
antiperiodic boundary conditions for their eigenstates,
respectively. 

The ``classical" trajectories are easily obtained by solving the
equations of motion that the bosonic and Grassmanian worldline 
actions in Eq.~(\ref{eq:Gfunct}) require. So, we arrive at
\begin{eqnarray}
&&x^{\mu}_{cl}(t)=\left(\frac{e^{2eFt}-1}{e^{2eF\tau}-1}
\right)^{\mu\nu}(z-y)_{\nu}+ \nonumber\\
&&+\int\limits_{0}^{\tau} dt' \left(
\frac{e^{2eFt}-1}{e^{2eF\tau}-1} 
\frac{ \left( e^{2eF(\tau-t')} -1 \right) }{eF}
- \theta(t-t') \frac{\left( e^{2eF(t-t')} -1 \right) }{eF}
\right)^{\mu\nu}\eta_{\nu}(t') ,
\label{eq:class}
\end{eqnarray}
and 
\begin{equation}
\psi^{\mu}_{cl}(t)=\int\limits_{0}^{\tau} dt' \left(
e^{2eF(t-t')}\left(\theta(t-t') -\frac{1}{1+e^{-2eF\tau}}\right)
\right)^{\mu\nu}\xi_{\nu}(t') .
\label{eq:classGr}
\end{equation}

Then, the result of the path integration in Eq.~(\ref{eq:Gfunct}) 
for the case of the coincident arguments $z=y=x$ reads
\begin{equation}
Z_{\tau}[\eta ,\xi ](x;x)=C_{0} \sqrt{\frac{Det(O_2)}{Det'(O_1)}}
\exp \left( \frac{i}{2} S^{bos}_{cl} [\eta ]
-\frac{1}{2} S^{fer}_{cl} [\xi ] \right) ,
\label{eq:Gfun}
\end{equation}
where the normalization constant $C_{0}$ should be determined 
by comparing the result with the Schwinger's one, or by satisfying
the normalization condition
\begin{equation}
Z_{\tau=0}[\eta ,\xi ](z;y)=\delta (z-y),
\end{equation}
which is equivalent to the operator equality $U(0)=1$. The prime in 
Eq.~(\ref{eq:Gfun}) denotes skipping a zero mode in the definition 
of the determinant. With our normalization convention for the 
determinants (see the next section), it is easy to check that the 
overall factor $C_{0}=-i/(2\pi\tau)^{2}$ in $3+1$ dimensions and
$C_{0}=\exp[-i\pi/4]/[2(\pi\tau)^{3/2}]$ in $2+1$ dimensions.

The expressions for $S^{bos}_{cl}$ and $S^{fer}_{cl}$ are
quadratic forms in the external sources
\begin{eqnarray} 
S^{bos}_{cl}[\eta]&=&\int\limits^{\tau}_{0} dt_{1} 
\int\limits^{\tau}_{0} dt_{2} \eta _{\nu }(t_{1}) 
D^{\nu}_{\lambda}(t_{1},t_{2}) \eta^{\lambda } (t_{2}),\\ 
S^{fer}_{cl}[\xi ] &=&
\int \limits^{\tau}_{0} dt_{1} \int \limits^{\tau}_{0} dt_{2} 
\xi_{\nu }(t_{1}) S^{\nu}_{\lambda }(t_{1},t_{2})
\xi^{\lambda}(t_{2}),
\end{eqnarray}
where the Green functions are given in terms of functions of 
the matrix argument $F_{\mu\nu}$
\begin{eqnarray} 
D(t_{1},t_{2})&=& \frac{1}{2eF} 
\Bigg[\epsilon (t_{1}-t_{2}) \left(1-e^{2eF(t_{1}-t_{2})}
\right)+\coth(eF\tau)\left(1 + e^{2eF(t_{1}-t_{2})} \right) 
\nonumber\\
&&-\frac{e^{eF (\tau-2t_{2})} + e^{eF(2t_{1}-\tau)}}
{\sinh (eF\tau)}\Bigg],
\label{eq:D}
\end{eqnarray}
\begin{equation}
S(t_{1},t_{2})= \frac{1}{2}\left[
\epsilon (t_{1}-t_{2}) - \tanh(eF \tau)
\right] e^{2eF (t_{1}-t_{2})}.
\label{eq:S}
\end{equation}

Substitution of Eqs. (\ref{eq:Gfun}), (\ref{eq:D}) and (\ref{eq:S}) 
into Eq.~(\ref{eq:trU2}) leads to the expression for $tr\langle 
x|U|x\rangle$. After expanding the exponent in powers of the 
operator valued interacting terms, $L_{bos}^{int}$ and
$L_{fer}^{int}$ (containing functional derivatives with respect to 
the sources $\eta_{\mu}(t)$ and $\xi_{\mu}(t)$), one has to
calculate the result of the derivative action on the generating
functional. Starting from this point, we have to restrict ourselves
to a specific finite number of the derivatives in the effective
action. As we mentioned before, in this paper we are interested 
in the two-derivative terms (see Sec.~\ref{Feynman} for some
discussions on computing the higher order approximations).
Therefore, we obtain
\begin{eqnarray} 
tr\langle x|U(\tau)|x\rangle&&=
\Bigg(1 +i\int\limits^{\tau}_{0} 
dt\left[V_{2}(t)+W_{2}(t)\right] -\frac{1}{2} \int 
\limits^{\tau}_{0} \int\limits^{\tau}_{0}dt_{1}dt_{2} 
\left[V_{1} (t_{1})V_{1}(t_{2})
+ W_{1}(t_{1})W_{1}(t_{2})\right] \nonumber\\ 
&& -\left. \int \limits^{\tau}_{0} 
\int\limits^{\tau}_{0}dt_{1}dt_{2} V_{1} (t_{1}) 
W_{1}(t_{2})\Bigg) 
Z_\tau[\eta ,\xi ](x,x) \right|_{\eta=0,\xi=0}, 
\label{eq:trU3} 
\end{eqnarray} 
where, as follows from Eqs. (\ref{eq:L2}), (\ref{eq:L3}) and 
(\ref{eq:trU2}), the vertex generating operators are
\begin{eqnarray} 
V_{1}(t)&=&\frac{i}{3} eF_{\nu\lambda,\mu}\lim_{t_{0}\to t} 
\frac{d}{dt_0} \frac{\delta^3}{\delta\eta _{\nu }(t_{0}) 
\delta\eta_{\lambda}(t) \delta\eta _{\mu }(t)}, \nonumber\\ 
V_{2}(t)&=&\frac{1}{8} eF_{\nu\lambda,\mu\kappa} 
\lim_{t_{0}\to t} \frac{d}{dt_0} 
\frac{\delta^4}{\delta\eta_{\nu}(t_{0}) \delta\eta_{\lambda}(t)
\delta\eta _{\mu}(t) \delta\eta_{\kappa}(t)}, \nonumber\\
W_{1}(t)&=&-eF_{\nu\lambda,\mu} \frac{\delta^{2}}
{\delta\xi_{\nu }(t) \delta\xi_{\lambda }(t)}\frac{\delta}
{\delta\eta_{\mu}(t)}, \nonumber\\ 
W_{2}(t)&=&\frac{i}{2}eF_{\nu\lambda,\mu\kappa}
\frac{\delta^{2}}{\delta\xi_{\nu }(t) \delta\xi_{\lambda}(t)}
\frac{\delta^2}{\delta\eta_{\mu}(t) \delta\eta_{\kappa}(t)}.
\end{eqnarray}
Substituting the generating functional (\ref{eq:Gfun}) which 
depends on the Green functions (\ref{eq:D}) and (\ref{eq:S}), we 
rewrite Eq.~(\ref{eq:trU3}) in the form 
\begin{eqnarray} 
&&tr\langle x|U(\tau)|x\rangle=
C_{0} \sqrt{ \frac{Det(O_2)} {Det'(O_1)}}
\Bigg\{ 1 -\frac{i}{8} eF_{\nu \lambda,\mu \kappa}
\int\limits_0^{\tau}dt \bigg[\dot{D}^{\nu\lambda }(t,t)
D^{\mu \kappa }(t,t)
\nonumber\\ &&+\dot{D}^{\nu \mu }(t,t) D^{\lambda \kappa }(t,t)
+\dot{D}^{\nu\kappa}(t,t) D^{\lambda\mu}(t,t) 
+4S^{\nu \lambda }(t,t)D^{\mu \kappa }(t,t)\bigg] \nonumber\\
&&-\frac{i}{18}e^{2} F_{\nu \lambda,\mu} F_{\sigma\kappa,\rho} 
\int\limits_0^{\tau}\int\limits_0^{\tau}dt_1 dt_2 \bigg[9D^{\mu 
\rho }(1,2) \left(S^{\kappa \sigma }(2,2)S^{\lambda \nu 
}(1,1)-2S^{\kappa \lambda }(2,1)S^{\sigma \nu }(2,1)\right)
\nonumber\\&&
+6S^{\sigma \kappa }(2,2)\left(\dot{D}^{\nu \lambda 
}(1,1)D^{\mu \rho }(1,2) +\dot{D}^{\nu \mu }(1,1)D^{\lambda \rho 
}(1,2)+\dot{D}^{\nu \rho }(1,2)D^{\lambda \mu }(1,1)\right)
\nonumber\\
&&+\dot{D}^{\nu \lambda }(1,1)\dot{D}^{\sigma \kappa 
}(2,2)D^{\mu \rho }(1,2) +2\dot{D}^{\nu \lambda 
}(1,1)\left(\dot{D}^{\sigma \rho }(2,2) D^{\mu \kappa }(1,2)
+\dot{D}^{\sigma \mu }(2,1)D^{\kappa \rho}(2,2)\right)
\nonumber\\
&&+\dot{D}^{\nu \mu 
}(1,1)\dot{D}^{\sigma \rho }(2,2)D^{\lambda \kappa }(1,2) 
+2\dot{D}^{\nu \kappa }(1,2)
\left(\dot{D}^{\sigma \rho }(2,2) D^{\lambda \mu }(1,1)
+\dot{D}^{\sigma \mu}(2,1)D^{\lambda \rho }(1,2)\right) 
\nonumber\\
&&+\dot{D}^{\nu \kappa }(1,2)\dot{D}^{\sigma \lambda }(2,1)
D^{\mu\rho}(1,2) +\dot{D}^{\nu \rho }(1,2) 
\dot{D}^{\sigma \mu }(2,1) D^{\lambda \kappa }(1,2) 
\nonumber\\
&&+\ddot{D}^{\nu \sigma }(1,2)\left(D^{\lambda \mu }(1,1)
D^{\kappa \rho } (2,2)+D^{\lambda \kappa }(1,2)
D^{\mu \rho}(1,2)+D^{\lambda \rho }(1,2) 
D^{\mu \kappa }(1,2)\right)\bigg]\Bigg\}.
\label{eq:tr-Gr}
\end{eqnarray}
Here the dotted functions are defined by the expressions
\begin{eqnarray}
\dot{D}^{\mu \nu}(1,2)&\stackrel{def}{=}&\frac{\partial}{\partial 
t_1} D^{\mu \nu}(t_1,t_2),\\
\ddot{D}^{\mu \nu}(1,2)&\stackrel{def}{=}&\frac{\partial ^2}
{\partial t_1\partial t_2}D^{\mu \nu}(t_1,t_2),\\
\dot{D}^{\mu \nu}(t,t)&\stackrel{def}{=}&\lim_{t_0\to 
t}\frac{\partial} {\partial t_0}D^{\mu \nu}(t_0,t).
\end{eqnarray}
Having the representation (\ref{eq:tr-Gr}) together with the Green 
functions (\ref{eq:D}) and (\ref{eq:S}), one is left with a need to 
perform the integrations over the proper time. This latter, however, 
may look like a rather complicated problem due to the necessity to 
disentangle the Lorentz indices while doing the integration. In the 
next section, we show how this problem can be solved.

\section{How to deal with functions of matrix argument 
$F_{\mu \nu }$} 
\label{Matr} 

In the previous section we developed the general method for
calculation the derivative expansion in QED. However, there 
was not given an explicit final expression, since we 
needed a technique dealing with functions of the matrix 
argument $F_{\mu\nu}$. Below we show, following the method of 
\cite{Batal}, how to deal with those functions as well as how 
to calculate the determinants of the differential operators in 
Eq.~(\ref{eq:operas}).

Let us begin by introducing notations that we are going to use 
below. When working with the electromagnetic field strength 
tensor, it is usually very convenient to introduce the 
invariants built of the field strength. In $(3+1)$-dimensional 
theory, the standard choice of the two independent invariants 
reads
\begin{eqnarray}
{\cal F}=-\frac{1}{4} F^{\mu \nu }F_{\mu \nu },\label{eq:F}
\qquad
{\cal G}=\frac{1}{8} \epsilon ^{\mu \nu \lambda \kappa } 
F_{\lambda \kappa } F_{\mu \nu } .
\end{eqnarray}
In our calculations, though, it will be more convenient to 
work with the following couple of invariants
\begin{eqnarray}
K_{+}=\sqrt{\sqrt{ {\cal F}^2+{\cal G}^2 }+{\cal F} },
\qquad
K_{-}=\sqrt{\sqrt{ {\cal F}^2+{\cal G}^2 }-{\cal F} }.
\end{eqnarray}
As for the $(2+1)$-dimensional theory, there exists only one 
independent invariant built of the electromagnetic field 
strength, and it is given by the expression analogues to 
${\cal F}$ in Eq.~(\ref{eq:F}).

Now we proceed to the case of $(3+1)$-dimensional QED. It is 
this case that was considered in \cite{Batal}. The authors of
that paper introduced the set of matrices $A^{\nu \lambda }_{(j)}$
with $j\in \{1,2,3,4\}$,
\begin{equation} 
A_{(j)\mu \nu}=\frac{-\bar{f}^{2}_{j} \eta_{\mu \nu } 
+ f_{j}F_{\mu \nu } + F^{2}_{\mu \nu} - 
i \bar{f}_{j} \stackrel{*}{F}_{\mu \nu} } 
{ 2 ( f^{2}_{j} - \bar{f}^{2}_{j} )}, \label{eq:matrA} 
\end{equation} 
where 
\begin{eqnarray} 
f_{1,2}= \pm i K_{-} &,\quad &   f_{3,4}= \pm  K_{+};\\ 
\bar{f}_{1,2}= \mp K_{+} &,\quad & \bar{f}_{3,4}= \mp i K_{-}. 
\end{eqnarray}

The main property of the matrices (\ref{eq:matrA}) that we are 
interested in are their (left and right) contractions with the 
field strength tensor,
\begin{equation}
F^{\nu \lambda }A_{(i)\lambda \mu} = A^{\nu \kappa}_{(i)}
F_{\kappa \mu} = f_{i}A^{\nu }_{(i)\mu }.
\label{eq:prop}
\end{equation}
Other useful properties of these matrices 
that will be used below are
\begin{eqnarray}
\sum_{j}A^{\mu \nu }_{(j)}=\eta^{\mu \nu },
\qquad A^{\mu }_{(j)\mu}=1,
\qquad A_{(k)}^{\mu \nu }A_{(j)\nu \lambda}=\delta_{kj}
A^{\mu }_{(j)\lambda}.&&
\label{eq:prop1}
\end{eqnarray}
As follows from the property in Eq.~(\ref{eq:prop}), for any 
function $\Phi(F)$ of the tensor argument $F_{\mu\nu}$, we get
\begin{equation} 
\Phi(F)_{\mu \nu} = \sum\limits_{j} A_{(j) \mu\nu} 
\Phi(f_{(j)}). 
\end{equation}

Matrices with similar properties can also be introduced for 
$(2+1)$-dimensional tensor $F_{\mu \nu }$ as well. Indeed, 
the following set of matrices
\begin{eqnarray} 
A^{\mu \nu}_{(\pm 1)}&=&\frac{1}{2}\left(\frac{(F^2)^{\mu \nu }}
{2{\cal F}} \pm \frac{F^{\mu \nu }}{\sqrt{2{\cal F}}} \right),
\qquad
A^{\mu \nu}_{(0)}= \eta^{\mu \nu }-\frac{(F^2)^{\mu \nu }}
{2{\cal F}}
\label{eq:matrA2+1} 
\end{eqnarray} 
in the $(2+1)$-dimensional case have properties similar to 
those in Eqs. (\ref{eq:prop}) and (\ref{eq:prop1}). As 
is easy to check directly, their eigenvalues are 
\begin{eqnarray} 
f_{\pm 1}= \pm \sqrt{2{\cal F}},\quad f_{0} = 0.
\end{eqnarray}

In particular, for the Green functions (\ref{eq:D}) and 
(\ref{eq:S}), which are functions of the tensor argument 
$F_{\mu\nu}$, we obtain the following representations, 
\begin{eqnarray} 
D^{\nu \lambda }(t_{1},t_{2})&=& 
\sum_{j}A^{\nu\lambda }_{(j)} 
\frac{1}{2ef_{j}} \Bigg[\epsilon 
(t_{1}-t_{2}) \left(1-e^{2ef_{j} (t_{1}-t_{2})}\right)
+\coth(ef_{j}\tau)\left(1 + e^{2ef_{j}(t_{1}-t_{2})} \right) 
\nonumber\\ &-& \frac{ e^{ef_{j} (\tau-2t_{2})} 
+ e^{ef_{j}(2t_{1}-\tau)}} { \sinh (ef_{j}\tau) }
\Bigg],
\label{eq:Dmunu}
\end{eqnarray}
\begin{equation}
S^{\nu \lambda }(t_{1},t_{2})= \sum_{j} 
A^{\nu \lambda }_{(j)} \frac{1}{2}
\left(\epsilon (t_{1}-t_{2}) 
-\tanh(ef_{j} \tau) \right) 
\exp[2ef_{j} (t_{1}-t_{2})].
\label{eq:Smunu}
\end{equation}
As is seen, in the case of vanishing field, the propagators 
$D^{\nu\lambda}(t_{1},t_{2})$ and $S^{\nu\lambda}(t_{1},t_{2})$ 
coincide with those used in \cite{McKeon,McK&Sher}.

Another problem is related to calculating the determinants of
the operators (\ref{eq:operas}). The latter are nothing else but 
products of all eigenvalues of the operators. Once again, making 
use of the matrices in Eq.~(\ref{eq:matrA}) or in
Eq.~(\ref{eq:matrA2+1}) for $(3+1)$- or $(2+1)$-dimensional cases, 
respectively, we look for the eigenvectors of the operators $O_1$ 
and $O_2$ in the form
\begin{eqnarray}
x^{\nu}_{(j)}(t)&=& A^{\nu}_{(j)\lambda} a^{\lambda } \phi(t)\\
\psi^{\nu}_{(j)}(t)&=& A^{\nu}_{(j)\lambda}\xi^{\lambda}\eta(t),
\end{eqnarray}
where $a^{\lambda } $ and $\xi^{\lambda } $ are constant nonzero 
vectors, $\phi$ and $\eta$ are scalar functions of $t$. As a 
result, the problem of obtaining eigenvalues reduces to solving 
ordinary differential equations for the scalar 
functions $\phi$ and $\eta$ with appropriate boundary conditions. 

Now, it is easy to check that, up to an unimportant constant, the 
corresponding determinants read (note that we skip a zero mode 
of the operator $O_1$)
\begin{eqnarray}
Det'^{(3+1)}(O_1)&=& \frac{\sinh^2(e\tau K_{+})}{(e\tau K_{+})^2} 
\frac{\sin^2(e\tau K_{-})}{(e\tau K_{-})^2}, \\
Det^{(3+1)}(O_2)&=&\cosh^2(e\tau K_{+})\cos^2(e\tau K_{-})
\end{eqnarray}
in the case of QED in $3+1$ dimensions, and 
\begin{eqnarray}
Det'^{(2+1)}(O_1)&=& \frac{\sinh^2(e\tau \sqrt{2{\cal F}})}
{(e\tau \sqrt{2{\cal F}})^2 }, \\
Det^{(2+1)}(O_2)&=&\cosh^2(e\tau \sqrt{2{\cal F}})
\end{eqnarray}
in the case of QED in $2+1$ dimensions. To obtain these results 
we used the following formulas for infinite products \cite{Ryzh}, 
\begin{eqnarray} 
\prod\limits_{n=1}^{\infty}\left(1+\frac{x^2}{\pi^2n^2}\right)=
\frac{\sinh x}{x}, &\qquad& 
\prod\limits_{n=1}^{\infty}
\left(1+\frac{4x^2}{\pi^2(2n+1)^2}\right)=\cosh x, 
\end{eqnarray}
and similar ones with replacement $x\to iy$.

\section{General Result in Spinor QED}
\label{GenerF} 

By making use of the results from the previous section, we can 
proceed with the calculation of (\ref{eq:tr-Gr}). 

After substituting the Green functions (\ref{eq:Dmunu}) and 
(\ref{eq:Smunu}), as well as the explicit expressions for the 
determinants of the operators $O_1$ and $O_2$, a straightforward,
though tedious computation gives the result for the diagonal matrix 
element of the $U(\tau)$,
\begin{eqnarray} 
&&tr\langle x|U(\tau)|x\rangle=tr\langle x|U(\tau)|x\rangle_{0}
\nonumber\\
&& \times\Bigg[1 -\frac{i}{8} eF_{\nu \lambda,\mu 
\kappa}\sum_{j,l} \bigg(C^{V}(f_{j},f_{l})\left(A^{\nu \lambda 
}_{(j)}A^{\mu \kappa }_{(l)}+ 2 A^{\nu \mu }_{(j)}A^{\lambda \kappa
\ }_{(l)}\right)+2C^{W}(f_{j},f_{l}) A^{\lambda \nu }_{(j)}A^{\mu 
\kappa }_{(l)}\bigg)\nonumber\\
&& -\frac{i}{18}e^{2}F_{\nu \lambda ,\mu } F_{\sigma 
\kappa, \rho} \sum_{j,l,k} \Bigg(9C^{WW}_{1}(f_{j},f_{l},f_{k}) 
A^{\kappa \sigma }_{(j)} A^{\lambda \nu }_{(l)}A^{\mu \rho 
}_{(k)}+9C^{WW}_{2}(f_{j},f_{l},f_{k}) A^{\kappa \lambda 
}_{(j)}A^{\sigma \nu }_{(l)}A^{\mu \rho }_{(k)}\nonumber\\
&&+ 6 C^{VW}_{1}(f_{j},f_{l},f_{k}) A^{\sigma \kappa }_{(j)} 
\left(A^{\nu \lambda}_{(l)}A^{\mu \rho }_{(k)}
+ A^{\nu \mu }_{(l)} A^{\lambda \rho}_{(k)}\right)
+6C^{VW}_{2}(f_{j},f_{l},f_{k}) A^{\sigma \kappa 
}_{(j)}A^{\nu \rho }_{(l)}A^{\lambda \mu }_{(k)}\nonumber\\
&& -C^{VV}_{1}(f_{j},f_{l},f_{k})\left(A^{\nu \lambda 
}_{(j)} A^{\kappa \sigma }_{(l)}A^{\mu \rho }_{(k)} + A^{\nu \mu 
}_{(j)}A^{\kappa \rho }_{(l)}A^{\lambda \sigma }_{(k)} +2 A^{\nu 
\lambda }_{(j)}A^{\kappa \rho }_{(l)}A^{\mu \sigma }_{(k)}\right) 
\nonumber\\
&& -C^{VV}_{2}(f_{j},f_{l},f_{k})\left(A^{\nu \sigma 
}_{(j)} A^{\kappa \lambda }_{(l)}A^{\mu \rho }_{(k)}+ A^{\nu \rho 
}_{(j)}A^{\kappa \mu }_{(l)}A^{\lambda \sigma }_{(k)}+ 2 A^{\nu 
\sigma }_{(j)}A^{\kappa \mu }_{(l)}A^{\lambda \rho }_{(k)}\right) 
\nonumber\\
&& - 2C^{VV}_{3}(f_{j},f_{l},f_{k}) \left(A^{\nu 
\lambda}_{(j)} A^{\kappa \mu }_{(l)}A^{\sigma \rho }_{(k)} + 
A^{\kappa \rho }_{(j)}A^{\nu \sigma }_{(l)}A^{\lambda \mu 
}_{(k)}\right) -C^{VV}_{4}(f_{j},f_{l},f_{k}) A^{\nu \kappa 
}_{(j)}A^{\lambda \mu }_{(l)} A^{\sigma \rho }_{(k)}\nonumber\\
&& -C^{VV}_{5}(f_{j},f_{l},f_{k}) A^{\nu \kappa }_{(j)} 
\left(A^{\lambda \sigma }_{(l)}A^{\mu \rho}_{(k)}+A^{\lambda \rho 
}_{(l)} A^{\mu \sigma}_{(k)}\right)\Bigg)\Bigg],
\label{eq:trUfer}
\end{eqnarray}
where the explicit expressions for the coefficients $C_{i}^{XY} 
(\alpha,\beta,\gamma)$ (with $X$, $Y\in \{V,W\}$) are given in 
Appendix~\ref{appA} and the diagonal matrix elements $tr\langle 
x|U(\tau)|x\rangle_{0}$ correspond to the non-derivative case,
\begin{eqnarray} 
tr\langle x|U(\tau)|x\rangle^{(3+1)}_{0} = 
- \frac{i} {4\pi ^{2}\tau^{2}} (e\tau K_{-}) (e\tau K_{+})
\cot(e\tau K_{-}) \coth(e\tau K_{+})
\end{eqnarray}
in $3+1$ dimensions, and
\begin{eqnarray}
tr\langle x|U_{0}(\tau)|x\rangle^{(2+1)}_{0} 
= \frac{\exp(-i\pi/4)}{2(\pi \tau)^{3/2}} 
(e\tau \sqrt{2{\cal F}})\coth(e\tau \sqrt{2{\cal F}}) 
\end{eqnarray}
in $2+1$ dimensions.

Equation (\ref{eq:trUfer}) (along with a similar one for scalar 
QED) is the main result of our paper. Note that the 
renormalization of the effective action (\ref{eq:ActGen3}) 
formally reduces to (i) performing a subtraction (precisely 
the same as in the original Schwinger's paper \cite{Sch}) of a 
term containing no derivatives of field strength with respect 
to coordinates, and (ii) changing all bare quantities for the 
renormalized ones, $e\to e_R$ and $A_{\mu} \to A_{\mu}^{R}$, 
defined as follows: 
\begin{equation}
 e_R=Z_3^{1/2}e,\qquad
A_{\mu}^{R}=Z_3^{-1/2}A_{\mu},\qquad Z_3^{-1}=1+{\bf C}e^2,
\label{eq:renorm}
\end{equation}
where
\begin{eqnarray}
{\bf C^{(3+1)}}&=&\frac{1}{12\pi^2} 
\int \limits^{\infty }_{1/\Lambda^2}
\frac{ds}{s} \exp(-sm^{2}),\label{ren3+1}\\
{\bf C^{(2+1)}}&=&\frac{1}{6\pi^{3/2}} 
\int \limits^{\infty }_{0}
\frac{ds}{\sqrt{s}} \exp(-sm^{2})=\frac{1}{6\pi m},
\label{ren2+1}
\end{eqnarray}
and $\Lambda$ is an ultraviolet cutoff in $(3+1)$-dimensional 
QED.

After subtraction and conversion to the renormalized quantities 
the effective action becomes finite in the limit 
$\Lambda\to\infty$. Since the derivative part of the effective 
action depends on $e$ and $A_{\mu}$ only through the product 
$eA_{\mu}=e_R A_{\mu}^{R}$ it does not change its form and no
further renormalization is required to make the derivative part 
well defined (below we use only renormalized quantities, 
although we always omit the script ``R" in their notation).
 
By using the asymptotic behavior of the coefficient functions 
(given in Appendix~\ref{appA}), one easily finds the following 
expansion of $tr\langle x|U(\tau)|x\rangle$ in powers of $\tau$,
\begin{eqnarray} 
&&tr\langle x|U(\tau)|x\rangle=tr\langle x|U(\tau)|x\rangle_{0}
\nonumber\\
&\times&\left[1 +\frac{ie^2\tau^3}{20} 
F^{\nu \lambda}F_{\nu\lambda,\mu}^{~~~~\mu}
+\frac{ie^2\tau^3}{180}\left( 
\frac{7}{2}F^{\nu \lambda,\mu}F_{\nu\lambda,\mu}-
F^{\nu \lambda,}_{~~~\lambda}F_{\nu\mu,}^{~~~\mu}
\right)+\dots\right].
\label{eq:trUferExp}
\end{eqnarray}
As is clear, this is the weak field limit of our general result 
in spinor QED. In the effective action, the given order in $\tau$ 
results in the two-derivative corrections of the order $1/m^2$:
\begin{equation} 
{\cal L}^{(3+1)spin}_{1/m^2}=\frac{\alpha}{720 \pi m^2} 
\left[18 F^{\nu \lambda}F_{\nu\lambda,\mu}^{~~~~\mu}
+7 F^{\nu \lambda,\mu}F_{\nu\lambda,\mu}-
2F^{\nu \lambda,}_{~~~\lambda}F_{\nu\mu,}^{~~~\mu}
\right],
\end{equation}
in $3+1$ dimensions, and of the order $1/m^3$
\begin{equation} 
{\cal L}^{(2+1)spin}_{1/m^3}=\frac{\alpha}{720 m^3} 
\left[18 F^{\nu \lambda}F_{\nu\lambda,\mu}^{~~~~\mu}
+7 F^{\nu \lambda,\mu}F_{\nu\lambda,\mu}-
2F^{\nu \lambda,}_{~~~\lambda}F_{\nu\mu,}^{~~~\mu}
\right],
\end{equation}
in $2+1$ dimensions.

The expansion in Eq.~(\ref{eq:trUferExp}) was obtained earlier 
in the heat kernel approach \cite{Hauk}. While the latter is a 
perfect tool for deriving the effective action in the weak field 
limit, it is not very useful when the field becomes strong. Our 
approach here, on the other hand, is free from such a limitation 
and the general result in Eq.~(\ref{eq:trUfer}) contains all 
the two derivative terms like $\partial F \partial F (F/m^2)^n$ 
where $n$ is an arbitrary positive integer and the Lorentz
indices (not shown) are contracted in all possible ways. To 
substantiate this claim, we present the next to leading terms 
of the weak field expansion in Eq.~(\ref{tau5fer}) in 
Appendix~\ref{appB}.

As we saw above, the formal expansion in $\tau$ corresponds to an 
expansion of the effective action in the inverse powers of the 
mass parameter. This means that, while making use of such an 
expansion, one cannot get any reliable results in the limit 
of the vanishing fermion mass. This, in particular, is the main 
reason why the authors of \cite{DitGi}, who used an expression 
like (\ref{eq:trUferExp}), came to a wrong conclusion about the 
absence of corrections to the one-loop effective action coming
from inhomogeneities of a static magnetic field when $m\to 0$. 
Such a conclusion ``contradicts" the result of Ref.~\cite{DHok}. 
The latter, as we will see, completely agrees with our result 
for the derivative expansion. 

\section{Spinor QED in 2+1 Dimensions}
\label{QED(2+1)F} 

Let us consider the case of the purely magnetic field background
to which a special attention was paid in \cite{DHok}. To proceed
with analyzing this case, note that the electromagnetic field
strength tensor takes the following form,

\begin{equation} 
F^{\mu \nu }(x)=B(x){\bf F}^{\mu \nu }, 
\label{eq:mag}
\end{equation} 
where $B(x)$ is a pseudoscalar function coinciding with the
magnetic field strength and ${\bf F}^{\mu \nu }$ is a constant
matrix with the only nonzero components ${\bf F}^{12}=-{\bf
F}^{21}=1$. As is seen it satisfies the following normalization
condition: ${\bf F}^{\mu \nu }{\bf F}_{\mu \nu }=2$. 

To reduce the general result presented in Eq.~(\ref{eq:trUfer})
for the particular choice of the field given in
Eq.~(\ref{eq:mag}), we have to use the properties of
$A_{(j)}^{\mu \nu}$'s presented in Sec.~\ref{Matr}. Just to get
feeling how they work, let us consider an example,
\begin{eqnarray} 
&&F_{\nu \lambda,\mu \kappa} \sum_{j,l} C^{W}(f_{j},f_{l})
A^{ \lambda \nu }_{(j)}A^{\mu \kappa }_{(l)}=
\frac{\partial_{\mu} \partial_{ \kappa } B }{B }
\sum_{j,l} C^{W}(f_{j},f_{l}) f_{(j)} A^{\mu \kappa }_{(l)}
\nonumber\\
&&=2\frac{\partial_{\mu} \partial_{ \kappa } B }{B }
\sqrt{2{\cal F}}\left[ C^{W}(\sqrt{2{\cal F}}, 0)
A_{(0)}^{\mu \kappa }+ 
C^{W}(\sqrt{2{\cal F}},\sqrt{2{\cal F}}) 
\left(A_{(-1)}^{\mu \kappa }+ A_{(+1)}^{\mu \kappa } \right)
\right]\nonumber \\
&&=-2iC^{W}(\sqrt{2{\cal F}},\sqrt{2{\cal F}}) 
\left({\bf F}^2\right)^{\mu \kappa } 
\partial_{\mu} \partial_{ \kappa } B 
=-2iC^{W}(\sqrt{2{\cal F}},\sqrt{2{\cal F}}) 
\sum\limits_{i=1}^{2}\partial_{i} \partial_{i} B .
\label{eq:exmpl}
\end{eqnarray} 
In this derivation, we made use of the Bianchi identity. We recall
that the latter should be satisfied since the electromagnetic field 
was introduced in the theory through the vector potential by minimal 
coupling. The identity itself reads $A_{(0)}^{\mu \nu} \partial_{\nu}
B \equiv 0$. The direct consequence of it is the independence of the 
magnetic field, for the particular choice (\ref{eq:mag}), on the time 
coordinate. By noticing that the matrix 
$A_{(0)}^{\mu \nu}$, as well as any other from the set, does not 
depend on $B(x)$ we obtain the secondary identity, 
$A_{(0)}^{\mu \nu} \partial_{\mu} \partial_{\nu} B \equiv 0$, 
by differentiating the original one. It is this last form of 
the Bianchi identity that was actually used in our derivation in 
Eq.~(\ref{eq:exmpl}). 

The other expressions, similar to that in Eq.~(\ref{eq:exmpl}), 
along with the functions like $C^{W}(\sqrt{2{\cal F}}, 
\sqrt{2{\cal F}})$ are listed in Appendix~\ref{appC}.

The final result for the derivative part of the diagonal matrix 
element (\ref{eq:trUfer}), for the particular choice of the field 
configuration in Eq.~(\ref{eq:mag}), reads
\begin{eqnarray} 
tr\langle x|U(\tau)|x\rangle_{der}^{(2+1)}&=& 
-\frac{ie^2 \left(\partial_{i}B\right)^2}
{(4\pi|eB|)^{3/2}} 
\frac{1}{\sqrt{\omega}} 
(3\omega ^{2} Y^{4} -3\omega Y^{3}-4\omega^{2}Y^{2}+3\omega Y
+\omega ^{2}) \nonumber \\ &=&
\frac{ie^2 \left(\partial_{i}B\right)^2}{(4\pi|eB|) ^{3/2}} 
\frac{\sqrt{\omega}}{2}\frac{d^3}{d\omega^3}
\left( \omega\coth\omega \right) ,
\label{eq:trUF2+1}
\end{eqnarray} 
where $\omega = i\tau |eB|$, $Y= \coth\omega$, and 
$\left(\partial_{i}B\right)^2\equiv\sum\limits_{i=1}^{2} 
\partial_{i} B \partial_{i} B$. Substituting the last expression 
into Eq.~(\ref{eq:ActGen3}), we come to the integral
representation for the derivative part of the effective
Lagrangian (we perform the change of the integration variable
$\tau$ for $\omega=i\tau |eB|$),
\begin{eqnarray} 
{\cal L}_{der}^{(2+1)spin}(B)&=&
-\frac{e^2 \left(\partial_{i}B\right)^2}{4(4\pi |eB|)^{3/2}} 
\int\limits_{0}^{\infty} \frac{d\omega}{\sqrt{\omega}} 
\exp \left( -\frac{m^2}{|eB|}\omega \right) 
\frac{d^3}{d\omega^3}\left( \omega \coth\omega \right) .
\label{eq:intrep}
\end{eqnarray} 
The last expression coincides with the result presented in
\cite{DHok} (note that in notation of \cite{DHok} $\partial_{i} B
\partial_{i} B = 4 \partial B\bar{\partial} B $). One can be 
convinced that the integrand in (\ref{eq:intrep}) is a negative function
what means that inhomogeneities of the magnetic field background, 
in approximation under consideration (one-loop and two derivatives), lead 
to the reduction of vacuum energy density for any value of the ratio 
$m^2/|eB|$. The latter situation does not, however, prove that a 
spontaneous generation of a non-homogeneous magnetic field happens in QED 
since the sign of the two derivative term in the expansion of the
effective action is not a sufficient argument for making a
conclusion of that kind \cite{DHok2}. 

We would like also to give another representation for the
derivative part of the Lagrangian in terms of special functions.
To get it, we need to perform the integration in
(\ref{eq:intrep}) by parts (see Eq.~(\ref{eq:d7}) in 
Appendix~\ref{appD}).
Here is such a representation,
\begin{eqnarray} 
{\cal L}_{der}^{(2+1)spin}(B)&=&-
\frac{e^2 \left(\partial_{i}B\right)^2} 
{\sqrt{2}\pi(4|eB|)^{3/2}} \Bigg[ 
5\zeta\left(-\frac{3}{2},1+\frac{m^2}{2|eB|}\right)
-9\frac{m^2}{2|eB|}\zeta\left(-\frac{1}{2},
1+\frac{m^2}{2|eB|}\right)\nonumber\\
&+&3\left(\frac{m^2}{2|eB|}\right)^2
\zeta\left(\frac{1}{2},1+\frac{m^2}{2|eB|}\right)
+\left(\frac{m^2}{2|eB|}\right)^3
\zeta\left(\frac{3}{2},1+\frac{m^2}{2|eB|}\right)
\Bigg].
\label{rep-zeta}
\end{eqnarray} 
Often, in the limit of large or small values of the external 
field, it is more convenient to work with the asymptotic 
expansions of the effective action rather than the exact 
expression as in Eq.~(\ref{rep-zeta}). First, let us consider
the case $m^2\ll |eB|$. Then, using the last representation, 
we easily derive the following asymptotic expansion, 
\begin{eqnarray} 
{\cal L}_{der}^{(2+1)spin}(B)&\simeq&
-\frac{e^2(\partial_{i}B)^2}{\sqrt{2}(4\pi|eB|)^{3/2}}
\sum\limits_{k=0}^\infty\frac{5-2k}{k!}
\Gamma\left(k+\frac{1}{2}\right)
\zeta\left(k-\frac{3}{2}\right)
\left(-\frac{m^2}{2|eB|}\right)^k.
\end{eqnarray} 
In order to get the asymptotic expansion for $m^2\gg |eB|$, 
we make use of the integral representation in Eq.~(\ref{eq:intrep})
and obtain
\begin{eqnarray} 
{\cal L}_{der}^{(2+1)spin}(B)&\simeq&
-\frac{e^2(\partial_iB)^2}{2\pi^{3/2}m^3}
\sum\limits_{k=0}^\infty\frac{B_{2k+4}}{(2k+1)!}
\Gamma\left(2k+\frac{3}{2}\right)
\left(\frac{2|eB|}{m^2}\right)^{2k},
\end{eqnarray} 
where $B_{k}$ are the Bernoulli numbers.

Now, let us consider the case of the purely electric field 
background. Without loosing the generality, we assume that the 
field is directed along the first axis of the two-dimensional 
space. Again the field strength tensor is factored similar to 
(\ref{eq:mag}),
\begin{equation} 
F^{\mu \nu }(x)=E(x){\bf F}^{\mu \nu }, 
\label{eq:elec}
\end{equation} 
where $E(x)$ is the magnitude of the electric field. Now the
constant matrix ${\bf F}^{\mu \nu }$ has nonzero components 
${\bf F}^{10} =-{\bf F}^{01}=1$, and satisfies the normalization
condition: ${\bf F}^{\mu\nu}{\bf F}_{\mu\nu}=-2$. The general
expression (\ref{eq:trUfer}) simplifies considerably for our 
choice of the background field. And the derivative part of that
expression now reads
\begin{eqnarray} 
tr\langle x|U(\tau)|x\rangle_{der}^{(2+1)}(E)&=& 
\frac{i\exp(-i\pi/4)}{(4\pi|eE|)^{3/2}}
\frac{e^2 \left(\partial_{\parallel}E\right)^2}{\sqrt{\omega}} 
(3\omega ^{2} Y^{4} -3\omega Y^{3}-4\omega^{2}Y^{2}+3\omega Y
+\omega ^{2}) \nonumber\\ 
&=&-\frac{i\exp(-i\pi/4)}{(4\pi|eE|)^{3/2}}
\frac{e^2 \left(\partial_{\parallel}E\right)^2}
{2}\sqrt{\omega}\frac{d^3}{d\omega^3}
\left( \omega \coth\omega \right) ,
\label{eq:trUF2+1E}
\end{eqnarray} 
where now $\omega = \tau |eE|$, $Y= \coth\omega$, and 
$\left(\partial_{\parallel}E\right)^2\equiv\left(\partial_{0}E 
\partial_{0}E- \partial_{1}E \partial_{1}E \right)$ . Here 
we used the Bianchi identity again to show that the electric field 
does not depend on the second spatial coordinate. Substituting 
this expression into Eq.~(\ref{eq:ActGen3}), we come to the 
integral representation for the derivative part of the effective 
Lagrangian,
\begin{eqnarray} 
{\cal L}_{der}^{(2+1)spin}(E)&=&
\frac{\exp(-i\pi/4) e^2 \left(\partial_{\parallel}E\right)^2}
{4(4\pi |eE|)^{3/2}} \int\limits_{0}^{\infty} 
\frac{d\omega}{\sqrt{\omega}} 
\exp \left( -i\frac{m^2}{|eE|}\omega \right) 
\frac{d^3}{d\omega^3}\left( \omega \coth\omega \right) .
\label{Im}
\end{eqnarray} 
As expected in the case of an electric field background, this
derivative correction to the effective action contains a nonzero
imaginary contribution. A convenient representation of the
latter can be obtained in the following way. First, in 
Eq.~(\ref{Im}), we switch to a new variable, $z=i\omega$, so 
that the integration runs along the imaginary axis of $z$ 
from zero to $i\infty$. Then, we move the 
integration contour to the real axis of $z$. As is easy to 
check, the integrand has poles at $z=\pi n$ ($n=1,2,\dots$). 
As a result, the real and the imaginary contributions get naturally 
separated. Indeed, the real part of ${\cal L}_{der}^{(2+1)spin}$ is 
given by the principal value of the integral along the $Re(z)$ 
axis, while the imaginary part appears due to the integration 
along the infinite set of the vanishingly small semi-circles 
above the poles, $z=\pi n +\varepsilon \exp[i(\pi-\phi)]$ 
(where $0<\phi<\pi$ and $\varepsilon\to 0$ at the end). In this 
way, we easily obtain the imaginary part of the right hand side 
in Eq.~(\ref{Im}),
\begin{eqnarray} 
{\cal I}m {\cal L}_{der}^{(2+1)spin}(E)&=& 
-\frac{e^2 \left(\partial_{\parallel}E\right)^2}
{2^8\pi^3 |eE|^{3/2}} \sum\limits_{n=1}^{\infty}
\frac{1}{n^{5/2}} \exp\left(-\frac{\pi m^2 n}{|eE|}\right)
\nonumber\\
&\times&\left[ 15 +18\frac{\pi m^2 n}{|eE|}
+12\left(\frac{\pi m^2 n}{|eE|}\right)^2
+8\left(\frac{\pi m^2 n}{|eE|}\right)^3
\right]. 
\label{eq:imag2+1} 
\end{eqnarray} 
We note that the result of the summation in the last expression
(as well as in similar formulas later on) can be given in terms
of the polylogarithmic function $\mbox{Li}_\nu (x)$ \cite{Lew}. 
Equation (\ref{eq:imag2+1}) determines the correction to the 
probability of the particle-antiparticle pair creation (by 
definition, the probability density is ${\cal W}=2{\cal I}m 
{\cal L}$) in an external electric field due to small
inhomogeneities in space-time. We emphasize that the correction
due to a time derivative of the field has the ``wrong" sign, i.e. 
it works against the particle creation. The gradient in the space 
direction parallel to the field strength, on the other hand, 
amplifies the process. 

As is known, in the case of constant electric field, the 
imaginary part of the effective Lagrangian is given by 
\begin{eqnarray} 
{\cal I}m {\cal L}^{(2+1)spin}(E)&=& 
\frac{|eE|^{3/2} }{4\pi^2} \sum\limits_{n=1}^{\infty}
\frac{1} {n^{3/2}}
\exp\left( -\frac{\pi m^2}{|eE|} n\right)
=\frac{|eE|^{3/2} }{4\pi^2} 
\mbox{Li}_{3/2}
\left[\exp\left(-\frac{\pi m^2}{|eE|} \right)\right]. 
\label{eq:imagF}
\end{eqnarray} 
This as well as the first correction due to the derivatives 
remain finite even in the limit of zero fermion mass. 
Despite of this fact, we still expect that the derivative 
expansion (with the electric field background) may fail in the 
limit of vanishingly small mass due to higher orders
in the number of derivatives. Below we shall see that the 
same is true in the spinor QED in $3+1$ dimensions as well.

\section{Spinor QED in 3+1 Dimensions}
\label{QED(3+1)F} 

As was mentioned at the beginning of the paper, the derivative 
expansion in QED$_4$ was also studied in \cite{Lee}. The result
of that paper was presented in an explicit form for the special
class of the electromagnetic field configurations, 
\begin{equation} 
{\cal G}=0,\qquad F^{\mu \nu }(x)=\Phi (x){\bf F}^{\mu \nu }, 
\label{eq:special}
\end{equation} 
where $\Phi (x)$ is a slowly varying function that defines the
magnitude of the field, and ${\bf F}^{\mu\nu}$ is a constant
matrix. For convenience, let us normalize the matrix ${\bf
F}^{\mu \nu }$ by the condition: ${\bf F}^{\mu \nu} {\bf F}_{\mu
\nu } = 2 $. Then the scalar function $\Phi (x)$ is nothing else
but $\sqrt{(-2{\cal F})}$. As was shown in our previous paper 
\cite{GS}, the general result for the diagonal matrix element 
(\ref{eq:trUfer}) in the case of field (\ref{eq:special}) reduces
to the same result as was presented in \cite{Lee},
\begin{eqnarray} 
tr\langle x|U(\tau)|x\rangle_{der}^{(3+1)}(\Phi)&=&\frac{1}{(4\pi 
)^{2}\tau} \frac{\partial _{\mu }\Phi \partial ^{\mu }\Phi }{\Phi 
^{2}} (3\omega ^{2} Y^{4} -3\omega Y^{3}-4\omega^{2}Y^{2}+3\omega 
Y+\omega ^{2}) , \nonumber\\
&=&-\frac{1}{(4\pi )^{2}\tau} \frac{\partial _{\mu }\Phi 
\partial ^{\mu }\Phi }{\Phi ^{2}} \frac{\omega}{2}
\frac{d^3}{d\omega^3}\left( \omega \coth\omega \right) , 
\label{eq:trUF}
\end{eqnarray} 
where $\omega = \tau e\Phi$, $Y= \coth\omega$. As in the 
$(2+1)$-dimensional theory, here we used the Bianchi
identity, which this time reads 
\begin{eqnarray} 
\left(\eta^{\mu\nu} + \left({\bf F}^2\right)^{\mu\nu}\right)
 \partial _{\nu }\Phi \equiv 0.
\end{eqnarray} 
In the case of magnetic field along the third axis, for example, 
this condition means that the specified field cannot depend on the 
time and the third spatial coordinates, while in the case of electric
field along the first axis, it cannot depend on the second and
third spatial coordinates.

Now, let us consider two particular cases of external 
field that we studied in $2+1$ dimensions: purely magnetic and 
purely electric field backgrounds. Both of them are just different 
possibilities of that given in Eq.~(\ref{eq:special}).

Thus, in the case of magnetic field (along the third axis in
space) we come to the following integral representation for the
derivative part of the effective Lagrangian,
\begin{eqnarray} 
{\cal L}_{der}^{(3+1)spin}(B)&=&
-\frac{e^2 \left(\partial_{i}B\right)^2}
{(8\pi)^{2}|eB|} 
\int\limits_{0}^{\infty} \frac{d\omega}{\omega} 
\exp \left( -\frac{m^2}{|eB|}\omega \right) 
\frac{d^3}{d\omega^3}\left( \omega \coth\omega \right) . 
\label{eq:inB3+1}
\end{eqnarray} 
Resembling the situation in $2+1$ dimensions, inhomogeneities of 
the external magnetic field tend to reduce vacuum energy 
density for any value of the ratio $m^2/|eB|$.

Performing integration in the right hand side of
Eq.~(\ref{eq:inB3+1}) by parts (see Eq.~(\ref{eq:d8}) in
Appendix~\ref{appD}), we find the following representation
(for the representation of the part of the effective action without
derivatives in terms of special functions, see \cite{Dittrich}),
\begin{eqnarray} 
{\cal L}_{der}^{(3+1)spin}(B)&=& 
-\frac{e^2 \left(\partial_{i}B\right)^2} {(8\pi)^2 |eB|}
\Bigg[ \frac{11}{6}\left(\frac{m^2}{|eB|}\right)^3
+\left(\frac{m^2}{|eB|}\right)^2 - \frac{1}{3}\frac{m^2}{|eB|}
-\left(\frac{m^2}{|eB|}\right)^3 \psi \left(1+\frac{m^2}{2|eB|}
\right) \nonumber\\
&+&24\zeta^{'}\left(-2,1+\frac{m^2}{2|eB|}\right)
-24\frac{m^2}{|eB|}\zeta^{'}\left(-1,1+\frac{m^2}{2|eB|}\right)
\nonumber\\
&+&6\left(\frac{m^2}{|eB|}\right)^2
\left[\ln\Gamma\left(1+\frac{m^2}{2|eB|}\right) - 
\ln\sqrt{2\pi}\right] \Bigg]. 
\label{eq:zetaB3+1}
\end{eqnarray} 
As $m^2\ll |eB|$, this expression allows the following 
asymptotic expansion,
\begin{eqnarray} 
{\cal L}_{der}^{(3+1)spin}(B)&\simeq& 
-\frac{e^2(\partial_iB)^2}{(8\pi)^2|eB|}
\Bigg[24\zeta'(-2)+\frac{2m^2}{3|eB|}-\frac{m^4}{2|eB|^2}
+\frac{m^6}{3|eB|^3}\nonumber \\
&&-\frac{m^8}{2|eB|^4}\sum_{k=0}^\infty\frac{k+1}{k+4}
\zeta(k+2)\left(-\frac{m^2}{2|eB|}\right)^k\Bigg],
\end{eqnarray}
where $\zeta'(-2)\approx -0.030$.
As $m^2\gg |eB|$, on the other hand, we obtain
\begin{eqnarray} 
{\cal L}_{der}^{(3+1)spin}(B)&\simeq&
-\frac{e^2(\partial_iB)^2}{(2\pi)^2m^2}\sum_{k=0}^\infty
\frac{B_{2k+4}}{2k+1}\left(\frac{2|eB|}{m^2}\right)^{2k}.
\end{eqnarray} 

In case of the electric field along the first axis, on the other
hand, we obtain the following expression for the derivative part 
of the effective action,
\begin{eqnarray} 
{\cal L}_{der}^{(3+1)spin}(E)&=&
-\frac{ie^2 \left(\partial_{\parallel}E\right)^2}
{(8\pi)^{2}|eE|} 
\int\limits_{0}^{\infty} \frac{d\omega}{\omega} 
\exp \left( -i\frac{m^2}{|eE|}\omega \right) 
\frac{d^3}{d\omega^3}\left( \omega \coth\omega \right) .
\label{eq:inE3+1}
\end{eqnarray} 
This expression has both real and imaginary part, as always
happens in the case of an external electric field. Another
representation for it is obtained by analytical continuation of
(\ref{eq:zetaB3+1}) according to the rule $|eB|\to -i |eE|$. The
imaginary part though is easily extracted from (\ref{eq:inE3+1})
in a standard way,
\begin{eqnarray} 
{\cal I}m {\cal L}_{der}^{(3+1)spin}(E)&=& 
\frac{e^2 \left(\partial_{\parallel}E\right)^2}{2^6\pi^4 |eE|} 
\sum\limits_{n=1}^{\infty}
\frac{1}{n^3} \exp\left(-\frac{\pi m^2 n}{|eE|}\right)
\nonumber\\ &\times& \left[6 +6\frac{\pi m^2 n}{|eE|}
+3\left(\frac{\pi m^2 n}{|eE|}\right)^2
+\left(\frac{\pi m^2 n}{|eE|}\right)^3
\right],
\label{eq:imag3+1}
\end{eqnarray} 
which determines a correction to the Schwinger result \cite{Sch} 
for the imaginary part of the effective action in a constant
electric field,
\begin{eqnarray} 
{\cal I}m {\cal L}^{(3+1)spin}(E)&=& 
\frac{(eE)^{2} }{8\pi^3} \sum\limits_{n=1}^{\infty}
\frac{1} {n^{2}}
\exp\left( -\frac{\pi m^2}{|eE|} n\right)
=\frac{(eE)^{2} }{8\pi^3}
\mbox{Li}_{2}
\left[\exp\left(-\frac{\pi m^2}{|eE|} \right)\right] .
\label{eq:imagin}
\end{eqnarray} 
The result in Eq.~(\ref{eq:imag3+1}) is in agreement with that of
\cite{DH}. 

As is easy to establish, both the Schwinger result for a
constant field and the first correction due to derivatives
are finite in the limit of the vanishing fermion mass. 
As we argued in the case of the $(2+1)$-dimensional spinor 
QED, this may not be the case in higher orders of the 
perturbative expansion in the number of derivatives.

\section{General Result in Scalar QED}
\label{GenerS} 

Now turning to the calculation of the derivative expansion for
the scalar electrodynamics, one does not need to repeat all the 
calculations similar to those done in Sec.~\ref{GenerF}. In order
to see this, we recall that the effective one-loop Lagrangian in
this case reads 
\begin{equation} 
L^{(1)scal}(x)=-i\int\limits^{\infty }_{0}\frac{d\tau}{\tau}
\langle x|U_{bos}(\tau)|x\rangle e^{-im^2\tau}.
\label{eq:Lbos}
\end{equation}
The evolution connected with the transition amplitude, $\langle 
z|U_{bos}(\tau)|y\rangle$, is described now by the Hamiltonian 
(compare with Eqs.~(\ref{eq:ActGen3}) and (\ref{eq:operH}))
\begin{equation}
H_{bos}=-\Pi_{\mu}\Pi^{\mu}, \quad 
\Pi_{\mu}=-i\partial _{ \mu } + eA_{\mu} (x).
\end{equation}
Thus, omitting all terms originating from the fermion part in 
the expression (\ref{eq:evol}), {\em i.e.} putting $L_{fer}^{int}=0$
in Eqs.~(\ref{eq:trU1}), (\ref{eq:trU2}) and $S^{fer}_{cl}=0$ in 
Eq.~(\ref{eq:Gfun}), we come to the following expression 
\begin{eqnarray} 
&&\langle x|U_{bos}(\tau)|x\rangle=
\langle x|U_{bos}(\tau)|x\rangle_{0}
\nonumber\\ 
&&\times \Bigg[1-\frac{i}{8}eF_{\nu \lambda ,\mu \kappa }\sum_{j,l} 
C^{V}(f_{j},f_{l})\bigg(A^{\nu \lambda }_{(j)}A^{\mu \kappa }_{(l)} 
+ 2 A^{\nu \mu }_{(j)}A^{\lambda \kappa \ }_{(l)}\bigg) 
+\frac{i}{18}e^{2}F_{\nu \lambda ,\mu } F_{\sigma 
\kappa,\rho}\nonumber\\
&&\times\sum_{j,l,k}\Bigg(C^{VV}_{1}(f_{j},f_{l},f_{k})\bigg(A^{\nu 
\lambda }_{(j)}A^{\kappa \sigma }_{(l)}A^{\mu \rho }_{(k)} + A^{\nu 
\mu }_{(j)}A^{\kappa \rho }_{(l)}A^{\lambda \sigma }_{(k)} +2 
A^{\nu \lambda }_{(j)}A^{\kappa \rho }_{(l)}A^{\mu \sigma 
}_{(k)}\bigg)\nonumber\\
&&+C^{VV}_{2}(f_{j},f_{l},f_{k}) \bigg(A^{\nu \sigma }_{(j)} 
A^{\kappa \lambda }_{(l)}A^{\mu \rho }_{(k)} +A^{\nu \rho 
}_{(j)}A^{\kappa \mu }_{(l)}A^{\lambda \sigma }_{(k)}+ 2 A^{\nu 
\sigma }_{(j)}A^{\kappa \mu }_{(l)}A^{\lambda \rho }_{(k)}\bigg) 
\nonumber\\
&&+ 2C^{VV}_{3}(f_{j},f_{l},f_{k}) \bigg(A^{\nu \lambda }_{(j)} 
A^{\kappa \mu }_{(l)}A^{\sigma \rho }_{(k)}+ A^{\kappa \rho }_{(j)} 
A^{\nu \sigma }_{(l)}A^{\lambda \mu }_{(k)}\bigg) 
+C^{VV}_{4}(f_{j},f_{l},f_{k})A^{\nu \kappa }_{(j)}A^{\lambda \mu 
}_{(l)} A^{\sigma \rho }_{(k)}\nonumber\\
&&+C^{VV}_{5}(f_{j},f_{l},f_{k}) A^{\nu \kappa }_{(j)} 
\bigg(A^{\lambda \sigma }_{(l)}A^{\mu \rho}_{(k)}+A^{\lambda \rho 
}_{(l)} A^{\mu \sigma}_{(k)}\bigg)\Bigg)\Bigg].
\label{eq:trUbos}
\end{eqnarray}
The coefficients used here are the same as in 
Eq.~(\ref{eq:trUfer}). As for the non-derivative factors, they have
the standard form,
\begin{eqnarray}
\langle x|U_{bos}(\tau)|x\rangle_{0}^{(3+1)} 
= -\frac{i}{(4\pi\tau)^{2}} 
\frac{(e\tau K_{-})(e\tau K_{+})} 
{\sin(e\tau K_{-})\sinh(e\tau K_{+})}
\end{eqnarray}
in $3+1$ dimensions, and 
\begin{eqnarray}
\langle x|U_{bos}(\tau)|x\rangle_{0}^{(2+1)} =
-\frac{ \exp(-i\pi/4) }{ (4\pi \tau)^{3/2} } 
\frac{(e\tau \sqrt{2{\cal F}})} 
{\sinh(e\tau\sqrt{2{\cal F}})}
\end{eqnarray}
in $2+1$ dimensions, as can be easily checked by using the 
expressions for the determinants given in Sec.~\ref{Matr} and by 
taking into account the fact that, because of spin degrees of 
freedom, we had the additional factor 4 for fermions.

In the case of scalar theory, the renormalization of the 
electromagnetic field and charge is given by the same 
formulas (\ref{eq:renorm}) but this time the corresponding 
constants read
\begin{eqnarray}
{\bf C^{(3+1)}}&=&\frac{1}{48\pi^2} 
\int \limits^{\infty }_{1/\Lambda^2}
\frac{ds}{s} \exp(-sm^{2}),\\
{\bf C^{(2+1)}}&=&\frac{1}{24\pi^{3/2}} \int \limits^{\infty }_{0}
\frac{ds}{\sqrt{s}} \exp(-sm^{2})=\frac{1}{24\pi m}.
\end{eqnarray}
 
To get a result of the type as in \cite{Hauk}, one has to expand 
the coefficient functions in powers of proper time. Thus the 
expansion for $\langle x|U_{bos}(\tau)|x\rangle$ (weak field 
limit) reads
\begin{eqnarray} 
&&\langle x|U_{bos}(\tau)|x\rangle=
\langle x|U_{bos}(\tau)|x\rangle_{0}
\nonumber\\
&\times& \left[1 -\frac{ie^2\tau^3}{30} 
F^{\nu \lambda}F_{\nu\lambda,\mu}^{~~~~\mu}
-\frac{ie^2\tau^3}{180}\left( 
4F^{\nu \lambda,\mu}F_{\nu\lambda,\mu}+
F^{\nu \lambda,}_{~~~\lambda}F_{\nu\mu,}^{~~~\mu}
\right)+\dots\right].
\label{eq:trUbosExp}
\end{eqnarray}
This expansion up to the order $\tau^5$ is given in 
Eq.~(\ref{tau5bos}) in Appendix~\ref{appB}. As in the spinor 
QED, it is useful only in the case of heavy scalar particles
(weak fields), when the mass scale is much larger than all 
other scales in the theory. 

In the effective action of scalar QED, the expansion in 
Eq.~(\ref{eq:trUbosExp}) corresponds to the following 
leading two derivative terms 
\begin{equation} 
{\cal L}^{(3+1)scal}_{1/m^2}=\frac{\alpha}{720 \pi m^2} 
\left[6 F^{\nu \lambda}F_{\nu\lambda,\mu}^{~~~~\mu}
+4F^{\nu \lambda,\mu}F_{\nu\lambda,\mu}+
F^{\nu \lambda,}_{~~~\lambda}F_{\nu\mu,}^{~~~\mu}
\right],
\end{equation}
in $3+1$ dimensions, and 
\begin{equation} 
{\cal L}^{(2+1)scal}_{1/m^3}=-\frac{\alpha}{720 m^3} 
\left[6 F^{\nu \lambda}F_{\nu\lambda,\mu}^{~~~~\mu}
+4F^{\nu \lambda,\mu}F_{\nu\lambda,\mu}+
F^{\nu \lambda,}_{~~~\lambda}F_{\nu\mu,}^{~~~\mu}
\right],
\end{equation}
in $2+1$ dimensions.

\section{Scalar QED in 2+1 Dimensions}
\label{QED(2+1)S} 

Let us start by considering the case of an external magnetic field 
as in Eq.~(\ref{eq:mag}). This time the derivative part of the
general expression (\ref{eq:trUbos}) reduces to 
\begin{eqnarray} 
\langle x|U_{bos}(\tau)|x\rangle_{der}^{(2+1)}&=& 
-\frac{ie^2 \left(\partial_{i}B\right)^2}{4(4\pi|eB|)^{3/2}} 
\frac{\sqrt{\omega}} {\sinh\omega}
(3 \omega Y^{3} - 3 Y^{2} - 2 \omega Y + 1) 
\nonumber\\ &=& 
\frac{ie^2 \left(\partial_{i}B\right)^2}{4(4\pi|eB|) ^{3/2}} 
\frac{\sqrt{\omega}}{2}
\left(\frac{d^3}{d\omega^3}+\frac{d}{d\omega}\right)
\left( \frac{\omega}{\sinh\omega} \right),
\label{eq:trUS2+1}
\end{eqnarray} 
where $\omega = i\tau |eB|$, $Y= \coth\omega$. After
substituting the last expression into Eq.~(\ref{eq:Lbos}), we 
come to the integral representation for the derivative part of the
effective Lagrangian (after performing the change of integration
variable $\tau\to \omega=i\tau |eB|$),
\begin{eqnarray} 
{\cal L}_{der}^{(2+1)scal}(B)&=& \frac{e^2
\left(\partial_{i}B\right)^2}{(16\pi |eB|)^{3/2}} 
\int\limits_{0}^{\infty} \frac{d\omega}{\sqrt{\omega}} 
\exp \left(-\frac{m^2}{|eB|}\omega \right) 
\left(\frac{d^3}{d\omega^3}+\frac{d}{d\omega}\right) \left(
\frac{\omega}{\sinh\omega} \right), 
\label{eq:L2+1B}
\end{eqnarray} 
which coincides with the result presented in \cite{DHok}. 
As in the case of spinor QED, there exists another 
representation of (\ref{eq:L2+1B}) given in terms of special 
functions (see Eq.~(\ref{eq:d14}) in Appendix~\ref{appD}),
\begin{eqnarray} 
&&{\cal L}_{der}^{(2+1)scal}(B)=
\frac{e^2 \left(\partial_{i}B\right)^2}
{\sqrt{2}\pi(16|eB|)^{3/2}} \Bigg[ 20
\zeta\left(-\frac{3}{2},\frac{1}{2}+\frac{m^2}{2|eB|}\right)
-18\frac{m^2}{|eB|}
\zeta\left(-\frac{1}{2},\frac{1}{2}+\frac{m^2}{2|eB|}\right)
\nonumber\\
&&+\left(1+3\left(\frac{m^2}{|eB|}\right)^2\right)
\zeta\left(\frac{1}{2},\frac{1}{2}+\frac{m^2}{2|eB|}\right)
+\frac{1}{2}\left(\frac{m^2}{|eB|}
+\left(\frac{m^2}{|eB|}\right)^3\right)
\zeta\left(\frac{3}{2},\frac{1}{2}+\frac{m^2}{2|eB|}\right)
\Bigg].
\end{eqnarray}
Numerical study of the integral in (\ref{eq:L2+1B}) shows that 
inhomogeneities of magnetic field background, in approximation 
under consideration (one-loop and two derivatives), lead to 
decreasing the vacuum energy density for $m^2/|eB| \agt 0.927$ 
and to increasing that density for $m^2/|eB| \alt 0.927$, in 
accordance with Ref.~\cite{DHok}. 

Analytically, we can obtain only the limiting cases as 
we did in spinor electrodynamics. In particular, for 
$m^2\ll |eB|$, the effective action takes the following 
asymptotic form,
\begin{eqnarray} 
{\cal L}_{der}^{(2+1)scal}(B)&\simeq& 
\frac{e^2(\partial_iB)^2}{(16\pi|eB|)^{3/2}}
\sum_{k=0}^\infty\frac{1}{k!}\Bigg[(2^k-2\sqrt{2})(5-2k)
\zeta\left(k-\frac{3}{2}\right)\nonumber\\
&&+\left(2^k-\frac{1}{\sqrt{2}}\right)(1-2k)
\zeta\left(k+\frac{1}{2}\right)\Bigg]
\Gamma\left(k+\frac{1}{2}\right)
\left(-\frac{m^2}{2|eB|}\right)^k, 
\end{eqnarray} 
while for $m^2\gg |eB|$, the expansion reads 
\begin{eqnarray} 
{\cal L}_{der}^{(2+1)scal}(B)&\simeq&
-\frac{e^2(\partial_iB)^2}{32\pi^{3/2}m^3}\sum_{k=0}^\infty
\frac{(2^{2k+3}-1)B_{2k+4}+(2^{2k+1}-1)B_{2k+2}}{(2k+1)!}
\nonumber\\
&\times&\Gamma\left(2k+\frac{3}{2}\right)
\left(\frac{|eB|}{m^2}\right)^{2k}.
\end{eqnarray} 

Now, let us consider the case of electric field background. 
Without loosing the generality, we assume that the field is 
directed along the first axis of space. We obtain 
\begin{eqnarray} 
\langle x|U_{bos}(\tau)|x\rangle_{der}^{(2+1)}(E)&=& 
\frac{i\exp(-i\pi/4)} {4(4\pi|eE|)^{3/2}} 
\frac{e^2 \left(\partial_{\parallel}E\right)^2\sqrt{\omega}}
{\sinh\omega} 
(3 \omega Y^{3} - 3 Y^{2} - 2 \omega Y + 1) 
\nonumber\\ &=&
-\frac{i\exp(-i\pi/4)}{(16\pi|eE|)^{3/2}}
e^2 \left(\partial_{\parallel}E\right)^2\sqrt{\omega}
\left(\frac{d^3}{d\omega^3}+\frac{d}{d\omega}\right)
\left( \frac{\omega}{\sinh\omega} \right),
\label{eq:trUS2+1E}
\end{eqnarray} 
where now $\omega = \tau |eE|$ and $Y= \coth\omega$. Substituting 
this expression into (\ref{eq:Lbos}), we come to the integral 
representation for the derivative part of the effective Lagrangian,
\begin{eqnarray} 
{\cal L}_{der}^{(2+1)scal}(E)&=&
-\frac{\exp(-i\pi/4) e^2 \left(\partial_{\parallel}E\right)^2}
{(16\pi |eE|)^{3/2}} \int\limits_{0}^{\infty} 
\frac{d\omega}{\sqrt{\omega}} 
\exp \left( -i\frac{m^2}{|eE|}\omega \right) 
\left(\frac{d^3}{d\omega^3}+\frac{d}{d\omega}\right)
\left( \frac{\omega}{\sinh\omega} \right).
\end{eqnarray} 
And we easily find the imaginary part of the expression,
\begin{eqnarray} 
&&{\cal I}m {\cal L}_{der}^{(2+1)scal}(E)=
\frac{e^2 \left(\partial_{\parallel}E\right)^2}
{2^9\pi^3 |eE|^{3/2}} 
\sum\limits_{n=1}^{\infty} \frac{(-1)^{n+1}}{n^{5/2}}
\exp\left(-\frac{\pi m^2 n}{|eE|}\right)\nonumber\\
&&\times\left[15 +18\frac{\pi m^2 n}{|eE|}
+4\pi^2 n^2 \left( 3\frac{m^4}{|eE|^2}-1\right)
+8\frac{m^2\pi^3 n^3}{|eE|}
\left(\frac{m^4}{|eE|^2} -1\right)
\right],
\label{eq:imag2+1S}
\end{eqnarray} 
which determines the correction to the corresponding result 
for case of constant electric field,
\begin{eqnarray} 
{\cal I}m {\cal L}^{(2+1)scal}(E)&=& 
\frac{|eE|^{3/2}}{8\pi^2}\sum\limits_{n=1}^{\infty}
\frac{(-1)^{n+1}} {n^{3/2}}
\exp\left( -\frac{\pi m^2}{|eE|} n\right)\nonumber\\
&=&-\frac{|eE|^{3/2}}{8\pi^2} \mbox{Li}_{3/2}
\left[-\exp\left(-\frac{\pi m^2}{|eE|} \right)\right] . 
\label{eq:imagS}
\end{eqnarray} 
A simple numerical analysis of the derivative correction in 
Eq.~(\ref{eq:imag2+1S}) shows that the sum in the right hand 
side, being positive for large values of the mass (or small 
values of the electric field), changes its sign at $m^2\approx 
0.721 |eE|$. Therefore, unlike the case of spinor QED, the 
time derivative of the field increases (while the gradient in 
space decreases) the probability of particle-antiparticle pair 
creation only for $m^2\agt 0.721 |eE|$. 

As in spinor QED, the two-derivative correction to the 
process of the pair production in scalar QED is convergent
even in the limit of the vanishing mass. This observation, of 
course, is not enough to prove that the derivative expansion 
is well defined to all orders in the massless theory. 

\section{Scalar QED in 3+1 Dimensions}
\label{QED(3+1)S} 

The derivative expansion for the electromagnetic field of the form 
(\ref{eq:special}) was presented in our previous paper \cite{GS}
(we just rewrite it in different form),
\begin{eqnarray} 
\langle x|U_{bos}(\tau)|x\rangle_{der}^{(3+1)}
&=&\frac{1}{(8\pi )^{2}\tau} 
\frac{\partial _{\mu }\Phi \partial ^{\mu }\Phi }{\Phi ^{2}} 
\frac{\omega}{\sinh\omega}
\left(3 \omega Y^3 - 3 Y^2 - 2 \omega Y + 1\right)\nonumber\\ 
&=&-\frac{1}{(8\pi )^{2}\tau} 
\frac{\partial _{\mu }\Phi \partial ^{\mu }\Phi }{\Phi ^{2}} 
\frac{\omega}{2}\left(\frac{d^3}{d\omega^3}+\frac{d}{d\omega}\right)
\left( \frac{\omega}{\sinh\omega} \right),
\label{eq:bosdiag}
\end{eqnarray}
with $\omega=\tau e \Phi$, $Y = \coth \omega$.

Now, let us consider the two most interesting particular cases as 
before. As in the case of the fermion theory presented in 
Sec.~\ref{QED(3+1)F}, in the case of scalar QED, the derivative
part of the effective Lagrangian is easily obtained by using 
(\ref{eq:bosdiag}) with $\Phi = i B$
\begin{eqnarray} 
{\cal L}_{der}^{(3+1)scal}(B)&=&
\frac{e^2 \left(\partial_{i}B\right)^2}{2(8\pi)^{2}|eB|} 
\int\limits_{0}^{\infty} \frac{d\omega}{\omega} 
\exp \left( -\frac{m^2}{|eB|}\omega \right) 
\left(\frac{d^3}{d\omega^3}+\frac{d}{d\omega}\right)
\left( \frac{\omega}{\sinh\omega} \right) .
\label{eq:inB3+1S}
\end{eqnarray} 
And again, as is easy to check, the situation with (\ref{eq:inB3+1S})
resembles that in $(2+1)$-dimensional scalar QED: inhomogeneities of 
the external magnetic field lead to decreasing the vacuum energy 
density for large values of the ratio $m^2/|eB|$ ($m^2/|eB|\agt 
0.41$) and to increasing for small values ($m^2/|eB|\alt 0.41$).

In addition to the representation (\ref{eq:inB3+1S}), we find the 
following one (see Eq.~(\ref{eq:d15}) in Appendix~\ref{appD})
\begin{eqnarray} 
{\cal L}_{der}^{(3+1)scal}(B)&=& 
\frac{e^2 \left(\partial_{i}B\right)^2} {2(8\pi)^2 |eB|}
\Bigg[ \frac{11}{6}\left(\frac{m^2}{|eB|}\right)^3
-\frac{m^2}{|eB|}\left(1+\left(\frac{m^2}{|eB|}\right)^2\right)
\psi\left(\frac{1}{2}+\frac{m^2}{2|eB|}\right)
\nonumber \\
&+&\frac{7}{6}\frac{m^2}{|eB|}
+2\left(1+3\left(\frac{m^2}{|eB|}\right)^2\right)
\left[\ln\Gamma\left(\frac{1}{2}+\frac{m^2}{2|eB|}\right) - 
\ln\sqrt{2\pi}\right]
\nonumber \\
&+&24\zeta^{'}\left(-2,\frac{1}{2}+\frac{m^2}{2|eB|}\right)
-24\frac{m^2}{|eB|}
\zeta^{'}\left(-1,\frac{1}{2}+\frac{m^2}{2|eB|}\right)
\Bigg].
\end{eqnarray} 
In the limit $m^2\ll |eB|$, this expression allows the following 
asymptotic expansion,
\begin{eqnarray} 
{\cal L}_{der}^{(3+1)scal}(B)&\simeq& 
\frac{e^2(\partial_iB)^2}{2(8\pi)^2|eB|}
\Bigg[-18\zeta'(-2)-\ln2+\frac{2m^2}{3|eB|}+\frac{m^6}{3|eB|^3}
\nonumber\\
&&-\frac{m^4}{2|eB|^2}\sum_{k=0}^\infty\frac{k+1}{k+2}
(2^{k+2}-1)\zeta(k+2)\left(-\frac{m^2}{2|eB|}\right)^k
\nonumber\\
&&-\frac{m^8}{2|eB|^4}\sum_{k=0}^\infty\frac{k+1}{k+4}
(2^{k+2}-1)\zeta(k+2)\left(-\frac{m^2}{2|eB|}\right)^k\Bigg]. 
\end{eqnarray} 
In the limit $m^2\gg |eB|$, on the other hand, we obtain 
\begin{eqnarray} 
{\cal L}_{der}^{(3+1)scal}(B)&\simeq&
-\frac{e^2(\partial_iB)^2}{(8\pi)^2m^2}\sum\limits_{k=0}^\infty
\frac{(2^{2k+3}-1)B_{2k+4}+(2^{2k+1}-1)B_{2k+2}}{2k+1}
\left(\frac{|eB|}{m^2}\right)^{2k}.
\end{eqnarray} 

In the case of the electric field directed along the 
first axis, we obtain
\begin{eqnarray} 
{\cal L}_{der}^{(3+1)scal}(E)&=&
\frac{ie^2 \left(\partial_{\parallel}E\right)^2}
{2(8\pi)^{2}|eE|} 
\int\limits_{0}^{\infty} \frac{d\omega}{\omega} 
\exp \left( -i\frac{m^2}{|eE|}\omega \right) 
\left(\frac{d^3}{d\omega^3}+\frac{d}{d\omega}\right) 
\left( \frac{\omega}{\sinh\omega} \right).
\label{eq:inE3+1S}
\end{eqnarray} 
Thus, the imaginary part of derivative part of the Lagrangian 
reads 
\begin{eqnarray} 
&&{\cal I}m {\cal L}_{der}^{(3+1)scal}(E)=
\frac{e^2 \left(\partial_{\parallel}E\right)^2}{2^7\pi^4 |eE|} 
\sum\limits_{n=1}^{\infty} \frac{(-1)^{n+1}}{n^3}
\exp\left(-\frac{\pi m^2 n}{|eE|}\right)\nonumber\\
&&\times\left[ 6 +6\frac{\pi m^2 n}{|eE|}
+\pi^2 n^2\left( 3\frac{m^4}{|eE|^2}-1\right)
+\frac{m^2\pi^3 n^3}{|eE|}\left(\frac{m^4}{|eE|^2}
-1\right) \right],
\label{eq:imag3+1S}
\end{eqnarray} 
which determines the correction to the probability of 
particle-antiparticle creation in a constant electric 
field expressed through
\begin{eqnarray} 
{\cal I}m {\cal L}^{(3+1)scal}(E)&=& 
\frac{(eE)^{2} }{16\pi^3}\sum\limits_{n=1}^{\infty}
\frac{(-1)^{n+1}} {n^{2}}
 \exp\left( -\frac{\pi m^2}{|eE|} n\right)\nonumber\\
&=&-\frac{(eE)^{2} }{16\pi^3} \mbox{Li}_{2}
\left[-\exp\left(-\frac{\pi m^2}{|eE|} \right)\right] .
\label{eq:imaginS}
\end{eqnarray} 
As in $(2+1)$-dimensional case, we observe that the sum in 
the right hand side of Eq.~(\ref{eq:imag3+1S}) is positive 
only for the large enough values of the mass ($m^2\agt 
0.388|eE|$). 

The expression (\ref{eq:imag3+1S}) concludes the list of our 
results describing the influence of slowly varying external 
electromagnetic fields on the spinor and scalar QED vacuum in 
two-derivative approximation. 

\section{How to get higher derivative terms?}
\label{Feynman}

Obviously, the method of the present paper can be applied for
calculating the higher derivative terms (with their total number
equal to four or higher) of the low energy effective action in
QED. However, the computational work with increasing the total
number of derivatives is getting so hard that obtaining already
all the four derivative terms seems to be impossible without use
of a computer. Just to get feeling how difficult this problem is,
let us consider the classification of all the relevant Feynman
diagrams in four derivative approximation.

To facilitate the calculation of the perturbative expansion in
number of derivatives in the problem at hand, it is appropriate
to develop the Feynman diagram technique. Our starting point will
be the system of equations (\ref{eq:trU2}) and (\ref{eq:Gfun}). 
Then, as is seen, the derivative expansion results from all 
(connected as well as disconnected) vacuum diagrams produced by 
(\ref{eq:trU2}). A somewhat disappointing feature of our
Lagrangian is an infinite number of local interactions.
Nevertheless, as will become clear in a moment, while working at
any finite order of the perturbative theory, one requires only a
finite number of those interactions. 

We observe that there are two different types of local
interactions in (\ref{eq:trU2}). The first (bosonic) type
contains only the bosonic fields, $x_{\mu}(t)$. The corresponding
vertices are shown in Figure~1. The other interactions involve
both the boson, $x_{\mu}(t)$, and the spinor fields,
$\psi_{\mu}(t)$.  These latter produce the vertices given in
Figure~2. The integers in the vertices denote the number of
derivatives (later called the weights of vertices) of the
electromagnetic field with respect to space-time. Some legs in
the diagrams are marked by circles and bullets. The circles
correspond to legs related to the first Lorentz index ($\nu$) of
the tensor weight, $F_{\nu\lambda,\mu_1,\dots,\mu_n}$, assigned
to the vertex, while the bullets, on the other hand, mark legs
which contain the derivatives with respect to the proper time. 
The latter act on the (bosonic) propagators attached to the
marked legs. 

The Feynman rules for writing expressions corresponding to Feynman 
diagrams are more or less standard. One has to use the propagators 
given in (\ref{eq:Dmunu}) and (\ref{eq:Smunu}) for connecting the 
bosonic (solid) and the fermion (dashed) legs, respectively. The 
combinatoric factors can be straightforwardly derived. Two simplest 
diagrams giving a nonzero contribution to two-derivative terms in 
the effective action are represented in Figure~3.

Let us mention the most general rules. To start with, we
classify all diagrams leading to terms with a given finite number
(called the weight of the corresponding diagrams from now on) of 
derivatives in the expansion. First of all, we see that diagrams 
with weight $N$ may contain different number of vertices. We 
denote by $Der(N)$ the set of all diagrams of a given weight $N$. 
By marking the bosonic (Fig.~1) and the fermion (Fig.~2) vertices 
with $n$ derivatives just by $[n]$ and $\overline{[n]}$, 
respectively, we see that the set $Der(N)$ contains a finite
number of elements: $Der(N)=\{\overline{[N]},[N], 
\overline{[N-1]}\oplus\overline{[1]}, \overline{[N-1]}\oplus[1], 
[N-1]\oplus\overline{[1]}, [N-1]\oplus[1],\dots\}$. Each of the 
elements in $Der(N)$ produces in its turn a (finite) number of 
Feynman diagrams differing from one another by all possible 
connections (by means of propagators) between all legs of the 
vertices. Thus, the diagrams of weight two in Figure~3, related
to $C^{W}$ and $C^{V}$ in the general expression
(\ref{eq:trUfer}), correspond to elements $\overline{[2]}$ and
$[2]$ in the set $Der(2)$, respectively.

Any element of $Der(N)$ specifies the number of different vertices 
as well as their separate weights. If the number of different
vertices in a diagram is given by integers
$\{V_1,V_2,\dots,V_k\}$ then the overall factor in front of the
corresponding expression is $1/(V_1!V_2!\dots V_k!)$. Next, let
the total number of bosonic and the fermion vertices are $k_B$
and $k_F$ (so that $k_B+k_F=k$), respectively. Then the total
number of bosonic legs of all the vertices in such a diagram is
$2k_B+N)$, while the number of the fermion legs is $2k_F$. Since,
we are interested in vacuum diagrams (with all legs being
connected) only, the diagrams of an odd weight $N$ are not
relevant for our derivative expansion. So, we put $N=2n$. As is
easy to count, the total number of all possible connections (by
means of $k_B+n$ bosonic and $k_F$ the fermion propagators)
between these vertices is $(N+2k_B-1)!!(2k_F-1)!!$, where we
assume that $(-1)!!\equiv 1$.  This is an upper bound for the
number of different diagrams with the given vertex set
corresponding to the given element $\overline{[V_1]}\oplus \dots
\oplus \overline{[V_{k_F}]} \oplus[V_{k_F+1}] \oplus \dots
\oplus[V_k] \in Der(N)$. However, due to the symmetry of the
vertices with respect to permutations of their non-marked legs as
well as with respect to permutations of identical vertices, some
of the diagrams are in fact equivalent.  For example, the naive
number of all relevant diagrams for the two-derivative terms in
the expansion of the effective action is $25$. On the other
hand, as is seen from our general result (\ref{eq:trUfer}), the
actual number of non-equivalent terms is $11$. 

Now let us say several words about the sign factors of diagrams. 
First, all diagrams of weight $N=2n$ have an overall factor 
$(-i)^n$. To get the right sign resulting from the the fermion
loops, one preliminary has to assign the direction of the the
fermion flow in the diagram by adding arrows on the the fermion
(dashed) lines. Then the overall sign factor is obtained by
multiplying sign factors for each the fermion loop of the
diagram. Each of the loop factors is defined by the formula: 
$(-1)^{N_0+1}$ where $N_0$ is the number of arrows running into
circles of loop vertices. This rule takes into account the fact
that the fermion propagators are antisymmetric with respect to
the simultaneous permutation of their Lorentz indices and proper
time coordinates as well as the fact that tensor weight at the
fermion vertices feels the order of first two indices.

Concluding this section, we would like to express a hope that the 
brief description of the Feynman technique given here would be 
enough for writing a code in some of the languages used for 
analytical computations if such a need appears.

\section{Conclusions}

In conclusion, here we further develop the method of our previous
paper \cite{GS} and generalize it to quantum electrodynamics in
$2+1$ dimensions. The distinctive feature of our approach is the
use of a special matrix basis (in Lorentz indices) in order to
deal with functions of antisymmetric tensors such as the 
(background) field strength tensor in QED. In Sec.~\ref{Matr}, we
give the explicit representation for these matrices as well as
demonstrate how they facilitate the calculation. It is also the
use of these matrices that allowed us to obtain the derivative
expansion in the fully covariant form.

Then, in this paper, we derived explicit expression of the
two-derivative term in the derivative expansion of the effective
action in QED in both fermion and scalar QED in $2+1$ and $3+1$
dimensions. In addition, we also calculated the leading order
corrections to the probability of the particle-antiparticle
creation rate produced by space-time gradients of the electric
field background. The latter gives a non-trivial generalization
of the famous Schwinger result in a constant electric field
\cite{Sch}.

Among other results, here we derived the Feynman rules for 
generating the perturbative expansion of the effective action in
the number of derivatives. This means that, in principle, an
arbitrary finite order of the derivative expansion is calculable
in our approach. For obvious reasons, the complexity of
calculation explodes at higher orders and, in the case of the
four-derivative approximation, the computational work already
becomes so hard that it is almost impossible to get a result in
the closed form without using a computer. By making use of the
Feynman rules, derived in this paper, one can write a computer
code in order to calculate higher order approximations.

At the end, let us also make a few remarks about possible tests
and applications of derivative expansion obtained in this paper.

As in the case of the Euler-Heisenberg action, the derivative
corrections will affect, among other things, the photon-photon
scattering amplitude. For a vanishing background field, the latter
is discussed in detail in \cite{Dicus}. Obviously, when the
background field is non-zero the corresponding amplitude and the
energy dependence of the cross section are going to change. As for
the explicit form of the result, it will be given elsewhere.

Besides that, it is likely that the explicit dependence of the
photon-photon cross section would be of great interest in studies
of some real systems which exist under extremely large magnetic
fields. The vicinity of the neutron stars and the early Universe
\cite{Olesen} are the most natural candidates of such systems.

The formal derivative expansion might also be useful in other
problems, such as the generalization of the theory of magnetic
catalysis of chiral symmetry breaking in $QED_4$ \cite{GMS} and
$QED_3$ \cite{Shpagin} to the case of inhomogeneous external
fields.

\section{Acknowledgments}
We would like to thank Theodore Hall for pointing out 
mistakes in the expressions for the imaginary part of the 
effective action in the first version of the paper.

V.P.G is grateful to the members of the Institute for Theoretical
Physics, University of Bern, Switzerland, for hospitality during his
stay there. The work of V.P.G. was supported by the grant
INTAS-93-2058-ext ``East-West network in constrained dynamical
systems'', by Swiss National Science Foundation grant
CEEC/NIS/96-98/7 IP 051219 and by Foundation of Fundamental
Researches of Ministry of Sciences of Ukraine under grant No.
2.5.1/003. The work of I.A.S. was supported by the U.S. Department 
of Energy Grant \#DE-FG02-84ER40153. 

\appendix

\section{Coefficient functions which appear in the derivative
expansion}
\label{appA}

Here we give the functions\footnote{Here we corrected the typos 
which appeared in \cite{GS}, namely we (i) omitted an extra term in
the expression for $C^{VV}_{4}$ that was mistakenly present; 
(ii) replaced the wrong factor $H(\tau\beta)$ in the last term of
$C^{VV}_{4}$ by $H(\tau\gamma)$, and (iii) added the third term in
$C^{VV}_{5}$ which was originally missing. In addition, we rewrote
$C^{VV}_{5}$ in a slightly different form.} used in
(\ref{eq:trUfer}) and (\ref{eq:trUbos}):  
\begin{eqnarray} 
C^{W}(\bar{\alpha},\bar{\beta} ) &=& \tau^{2}\tanh(\alpha 
\tau)H(\beta \tau) ,\\ 
C^{V}(\bar{\alpha},\bar{\beta} ) &=& \alpha \tau^{3}H(\alpha 
\tau)H(\beta \tau)- \frac{\alpha \tau}{\beta ^{2}-\alpha 
^{2}}[H(\beta \tau)-H(\alpha \tau)],\\ 
C^{WW}_{1}(\bar{\alpha},\bar{\beta},\bar{\gamma} ) 
&=&\frac{\tau^{3}}{8} \tanh(\alpha \tau)\tanh(\beta \tau)H(\gamma 
\tau),\\ 
C^{WW}_{2}(\bar{\alpha},\bar{\beta},\bar{\gamma}) 
&=&\frac{\tau^{2}}{4}[\tanh(\alpha \tau)+ \tanh(\beta 
\tau)]\left(\frac{H(\alpha \tau+\beta \tau)-H(\gamma \tau)} {\alpha 
+\beta -\gamma }-\frac{H(\gamma \tau)}{\alpha +\beta }\right),\\ 
C^{VW}_{1}(\bar{\alpha},\bar{\beta},\bar{\gamma}) &=& 
-\frac{\tau^{3}}{4} \tanh(\alpha \tau)\left(\beta \tau H(\beta 
\tau)H(\gamma \tau)-\frac{H(\beta \tau)- H(\gamma \tau)}{\tau(\beta 
+\gamma )}\right) ,\\ 
C^{VW}_{2}(\bar{\alpha},\bar{\beta},\bar{\gamma}) &=& 
\frac{\tau^{2}\beta \tanh(\alpha \tau)} {2(\beta ^{2}-\gamma ^{2})} 
\left[H(\beta \tau)-H(\gamma \tau)\right] , 
\end{eqnarray} 
\begin{eqnarray} 
C^{VV}_{1}(\bar{\alpha},\bar{\beta},\bar{\gamma}) 
&=&\frac{\tau^{5}\alpha \beta}{2}H(\alpha \tau)H(\beta 
\tau)H(\gamma \tau) -\frac{\tau^{3}\alpha H(\alpha \tau)}{2(\beta 
-\gamma )} \left(H(\beta \tau)-H(\gamma \tau)\right)\nonumber\\ 
&-&\frac{\tau^{3}\beta H(\beta \tau)}{2(\alpha +\gamma )} 
\left(H(\alpha \tau)-H(\gamma \tau)\right)-\frac{\tau}{2} 
\frac{H(\alpha \tau)}{(\alpha +\gamma )(\alpha +\beta )}\nonumber\\ 
&-&\frac{\tau}{2} \frac{H(\beta \tau)}{(\alpha +\beta )(\beta 
-\gamma )} + \frac{\tau}{2} \frac{H(\gamma \tau)}{(\alpha +\gamma 
)(\beta -\gamma )} ,\\ 
C^{VV}_{2}(\bar{\alpha},\bar{\beta},\bar{\gamma}) &=& 
-\frac{\tau^{3} \alpha \beta H(\alpha \tau)H(\beta \tau)}{2(\alpha 
-\beta )(\alpha -\beta +\gamma )}+ \frac{\tau^{3}[2(\alpha -\beta 
)+\gamma \ ]H(\gamma \tau)\left[\beta H(\beta \tau)-\alpha H(\alpha 
\tau)\right]}{2(\alpha -\beta ) (\alpha -\beta +\gamma) 
}\nonumber\\ &+&\frac{\alpha \tau}{2}H(\alpha 
\tau)\left(\frac{2(\beta +\gamma )} {(\alpha ^{2}-\beta 
^{2})(\alpha ^{2}-\gamma ^{2})} - \frac{2\alpha - \beta 
+\gamma}{(\alpha -\beta )^{2}(\alpha +\gamma )(\alpha -\beta 
+\gamma )} \right)\nonumber\\ &+&\frac{\beta \tau}{2}H(\beta 
\tau)\left(\frac{2(\gamma -\alpha )} {(\alpha ^{2}-\beta 
^{2})(\beta ^{2}-\gamma ^{2})}+ \frac{2\beta -\alpha - 
\gamma}{(\alpha -\beta )^{2}(\beta -\gamma )(\alpha -\beta +\gamma 
)}\right)\nonumber\\ &+&\frac{\tau}{2}H(\gamma 
\tau)\left(\frac{2(\alpha \beta +\gamma ^{2})} {(\alpha ^{2}-\gamma 
^{2})(\beta ^{2}-\gamma ^{2})} - \frac{\gamma} {(\alpha +\gamma 
)(\beta -\gamma )(\alpha -\beta +\gamma )}\right)\nonumber\\ 
&+&\frac{\tau}{2(\alpha -\beta )(\alpha -\beta +\gamma )},\\ 
C^{VV}_{3}(\bar{\alpha},\bar{\beta},\bar{\gamma}) &=& 
-\frac{\tau^{3}\alpha \beta H(\alpha \tau)}{\beta ^{2}-\gamma ^{2}} 
\left[H(\beta \tau)-H(\gamma \tau)\right] +\frac{\alpha \tau 
H(\alpha \tau)}{(\alpha -\beta )(\alpha ^{2}- \gamma 
^{2})}\nonumber\\ &-&\frac{\tau}{\beta ^{2}-\gamma ^{2}} 
\left(\frac{\beta}{\alpha -\beta } H(\beta \tau)-\frac{\alpha \beta 
+\gamma ^{2}}{\alpha ^{2}-\gamma ^{2}} H(\gamma \tau) \right) ,\\ 
C^{VV}_{4}(\bar{\alpha},\bar{\beta},\bar{\gamma}) &=& 
-\frac{2\tau\alpha^{2}H(\alpha \tau)} {(\alpha ^{2}-\beta 
^{2})(\alpha ^{2}- \gamma ^{2})} +\frac{2\tau\beta ^{2}H( \beta 
\tau)} {(\alpha ^{2}-\beta ^{2})(\beta ^{2}-\gamma ^{2})} 
+\frac{2\tau\gamma ^{2}H(\gamma \tau)} {(\alpha 
^{2}-\gamma^{2})(\gamma^{2}- \beta ^{2})},\\ 
C^{VV}_{5}(\bar{\alpha},\bar{\beta},\bar{\gamma}) &=&\frac{\tau^3 
\alpha }{2} \left(\frac{\alpha}{(\alpha +\beta )(\alpha +\beta 
+\gamma )}+\frac{1} {\alpha +\gamma }\right) H(\alpha \tau)H(\beta 
\tau)\nonumber\\ &+&\frac{\tau^3 \alpha }{2} 
\left(\frac{\alpha}{(\alpha +\gamma ) (\alpha +\beta +\gamma) 
}+\frac{1}{\alpha +\beta }\right) H(\alpha \tau) H(\gamma 
\tau)\nonumber\\ &+&\tau^3H(\beta \tau)H(\gamma \tau) 
+\frac{\tau^{3}}{2}\left(\frac{\beta \gamma (\beta +\gamma 
)}{\alpha + \beta +\gamma }-2\alpha ^{2}\right) \frac{H(\beta 
\tau)H(\gamma \tau)} {(\alpha +\beta )(\alpha +\gamma )}\nonumber\\ 
&+&\frac{\alpha \tau H(\alpha \tau) \left(2+\frac{\alpha 
+\beta}{\alpha +\gamma }+\frac{\alpha +\gamma}{\alpha + \beta 
}\right) }{ 2(\alpha +\beta )(\alpha +\gamma )(\alpha +\beta 
+\gamma )} +\frac{\tau}{2}\Bigg(\frac{2H(\gamma \tau)} 
{(\alpha +\gamma )(\beta -\gamma)} \nonumber\\ 
&-& \frac{2H(\beta \tau)} {(\alpha +\beta )(\beta -\gamma)} 
+\frac{\gamma H(\gamma \tau)}{(\alpha +\gamma )^{2} 
(\alpha +\beta +\gamma )}
+\frac{\beta H(\beta \tau)}{(\alpha +\beta)^{2} 
(\alpha +\beta +\gamma )} \nonumber\\ 
&-& \frac{1}{(\alpha +\beta) (\alpha +\beta +\gamma) }- 
\frac{1}{(\alpha +\gamma )(\alpha +\beta +\gamma )} \Bigg). 
\end{eqnarray} Here we used the following notation 
\begin{eqnarray} 
H(x)&=&\frac{x\coth{x}-1}{x^{2}}, 
\end{eqnarray} 
and the letters with bars differ from the letters without those 
only in a factor of the electric charge: $\alpha=e\bar{\alpha}$. 
Note that in \cite{GS} we ignored this difference.

As $\tau \to 0$, these coefficient functions have the following 
asymptotic behavior
\begin{eqnarray} 
C^{W}(\bar{\alpha},\bar{\beta} ) &\simeq& 
\frac{\alpha\tau^{3}}{3}
-\frac{\alpha\tau^{5}}{45}\left(5\alpha^2+\beta^2\right)
+O\left(\tau^7\right) ,\\ 
C^{V}(\bar{\alpha},\bar{\beta} ) &\simeq& 
\frac{2\alpha\tau^{3}}{15}
-\frac{\alpha\tau^{5}}{105}\left(\alpha^2+\beta^2\right)
+O\left(\tau^7\right) ,\\ 
C^{WW}_{1}(\bar{\alpha},\bar{\beta},\bar{\gamma} ) 
&\simeq& 
\frac{\alpha\beta\tau^{5}}{24}
+O\left(\tau^7\right) ,\\ 
C^{WW}_{2}(\bar{\alpha},\bar{\beta},\bar{\gamma} )&\simeq& 
-\frac{\tau^{3}}{12}
+\frac{\tau^{5}}{180}\left(4\alpha^2+4\beta^2+\gamma^2
-7\alpha\beta-\alpha\gamma-\beta\gamma\right)
+O\left(\tau^7\right) ,\\ 
C^{VW}_{1}(\bar{\alpha},\bar{\beta},\bar{\gamma} ) &\simeq& 
\frac{\alpha\tau^{5}}{180}\left(\gamma-6\beta\right)
+O\left(\tau^7\right) ,\\ 
C^{VW}_{2}(\bar{\alpha},\bar{\beta},\bar{\gamma} )&\simeq& 
-\frac{\alpha\beta\tau^{5}}{90}
+O\left(\tau^7\right) ,\\ 
C^{VV}_{1}(\bar{\alpha},\bar{\beta},\bar{\gamma} ) 
&\simeq&\frac{\tau^{3}}{90}
-\frac{\tau^{5}}{1890}\left(2\alpha^2+2\beta^2+2\gamma^2
-51\alpha\beta-9\alpha\gamma+9\beta\gamma\right)
+O\left(\tau^7\right) ,\\ 
C^{VV}_{2}(\bar{\alpha},\bar{\beta},\bar{\gamma} ) 
&\simeq&-\frac{4\tau^{3}}{45}
+\frac{\tau^{5}}{1890}\left(10\alpha^2+10\beta^2+13\gamma^2
+15\alpha\beta+3\alpha\gamma-3\beta\gamma\right)
+O\left(\tau^7\right) ,\\ 
C^{VV}_{3}(\bar{\alpha},\bar{\beta},\bar{\gamma} ) 
&\simeq&-\frac{\tau^{3}}{45}
+\frac{\tau^{5}}{945}\left(2\alpha^2+2\beta^2+2\gamma^2
+9\alpha\beta\right)
+O\left(\tau^7\right) ,\\ 
C^{VV}_{4}(\bar{\alpha},\bar{\beta},\bar{\gamma} ) 
&\simeq&\frac{2\tau^{3}}{45}
-\frac{4\tau^{5}}{945}\left(\alpha^2+\beta^2+\gamma^2\right)
+O\left(\tau^7\right) ,\\ 
C^{VV}_{5}(\bar{\alpha},\bar{\beta},\bar{\gamma} ) 
&\simeq&\frac{8\tau^{3}}{45}
-\frac{\tau^{5}}{1890}\left(20\alpha^2+23\beta^2+23\gamma^2
-12\alpha\beta-12\alpha\gamma+6\beta\gamma\right)
+O\left(\tau^7\right) .
\end{eqnarray} 

\section{Expansion of the derivative terms in powers of 
the proper time}
\label{appB}

In this appendix we give the proper time expansion of the 
derivative terms, as in Eqs.~(\ref{eq:trUferExp}) and 
(\ref{eq:trUbosExp}), up to the order $\tau^5$. 

In case of spinor QED, from Eq.~(\ref{eq:trUfer}) we derive 
the expansion 
\begin{eqnarray} 
&&tr\langle x|U(\tau)|x\rangle \simeq 
tr\langle x|U(\tau)|x\rangle_{0} 
\Bigg[1 +\frac{ie^2\tau^3}{20} 
F^{\nu \lambda}F_{\nu\lambda,\mu}^{~~~~\mu}
\nonumber\\
&&+\frac{ie^2\tau^3}{180}\left( 
\frac{7}{2}F^{\nu \lambda,\mu}F_{\nu\lambda,\mu}-
F^{\nu \lambda,}_{~~~\lambda}F_{\nu\mu,}^{~~~\mu}
\right)
-\frac{ie^4\tau^5}{315} F_{\nu\lambda,\mu\kappa}
\left(16 {\cal F} \eta^{\mu \kappa} F^{\nu \lambda} 
  + F^{\nu \lambda} (F^{2})^{\mu \kappa}
\right)\nonumber\\
&&+\frac{ie^4\tau^5}{1890} \Bigg(
2 F_{\nu\lambda,\mu} F^{\nu\rho,}_{~~~\rho} (F^{2})^{\lambda \mu } 
-2 F_{\nu\lambda,\mu}F^{\nu\lambda,\rho} (F^{2})^{\mu}_{~\rho} 
-37 F_{\nu\lambda,\mu}F^{\nu\sigma,\mu}(F^{2})^{\lambda}_{~\sigma}
+ F_{\nu\mu,}^{~~~\mu} F_{\sigma\rho,}^{~~~\rho} (F^{2})^{\nu
\sigma} \Bigg)\nonumber\\
&&-\frac{ie^4\tau^5}{2520}
  F_{\nu\lambda,\mu} F_{\sigma\kappa,\rho} \Bigg(
 38 \eta^{\mu \rho} F^{\lambda \sigma} F^{\nu \kappa} 
-12 \eta^{\nu \kappa} F^{\lambda \rho} F^{\mu \sigma} 
+47 \eta^{\mu \rho} F^{\nu \lambda} F^{\sigma \kappa} 
+16 \eta^{\kappa \rho} F^{\lambda \sigma} F^{\nu \mu } 
\Bigg)\Bigg].
\label{tau5fer}
\end{eqnarray}
Notice that despite the difference between the 
two sets of matrices $A_{(j)}^{\mu\nu}$ in $2+1$ and $3+1$ 
dimensions, the expression in square brackets is independent 
of the dimension up to this order in the expansion. In calculation,
we took into account the Bianchi identity to show that many
seemingly different terms appearing in the expansion reduce to the 
same structures. In particular, the following relations are 
the identities that we needed
\begin{eqnarray}
F_{\nu\lambda,\mu\kappa} \eta^{\lambda \kappa} F^{\nu \mu} 
&=&\frac{1}{2}F_{\nu\lambda,\mu\kappa} \eta^{\mu \kappa} 
F^{\nu \lambda}, \\
F_{\nu\lambda,\mu\kappa} F^{\nu \mu} (F^{2})^{\lambda \kappa} 
&=&\frac{1}{2}F_{\nu\lambda,\mu\kappa} F^{\nu \lambda} 
(F^{2})^{\mu \kappa} ,\\
 F_{\nu\lambda,\mu} F_{\sigma\kappa,\rho}
\eta^{\nu \rho} F^{\lambda \sigma} F^{\mu \kappa} 
&=& \frac{1}{2}F_{\nu\lambda,\mu} F_{\sigma\kappa,\rho}
\eta^{\mu \rho} F^{\lambda \sigma} F^{\nu \kappa} ,\\
F_{\nu\lambda,\mu} F_{\sigma\kappa,\rho}
\eta^{\nu \kappa} F^{\lambda \sigma} F^{\mu \rho} 
&=& F_{\nu\lambda,\mu} F_{\sigma\kappa,\rho}
 \eta^{\nu \kappa} F^{\lambda \rho} F^{\mu \sigma} 
+ \frac{1}{2}F_{\nu\lambda,\mu} F_{\sigma\kappa,\rho}
\eta^{\mu \rho} F^{\lambda \sigma} F^{\nu \kappa},\\
F_{\nu\lambda,\mu} F_{\sigma\kappa,\rho}
\eta^{\mu \sigma} F^{\kappa \rho} F^{\nu \lambda} 
&=&-\frac{1}{2}F_{\nu\lambda,\mu} F_{\sigma\kappa,\rho}
\eta^{\mu \rho} F^{\nu \lambda} F^{\sigma \kappa},\\
F_{\nu\lambda,\mu} F_{\sigma\kappa,\rho}
\eta^{\sigma \rho} F^{\mu \kappa} F^{\nu \lambda} 
&=&-2F_{\nu\lambda,\mu} F_{\sigma\kappa,\rho}
\eta^{\kappa \rho} F^{\lambda \sigma} F^{\nu \mu } ,\\
F_{\nu\lambda,\mu} F_{\sigma\kappa,\rho}
\eta^{\lambda \sigma} F^{\kappa \rho} F^{\nu \mu } 
&=&-\frac{1}{4}F_{\nu\lambda,\mu} F_{\sigma\kappa,\rho}
\eta^{\mu \rho} F^{\nu \lambda} F^{\sigma \kappa},\\
F_{\nu\lambda,\mu} F_{\sigma\kappa,\rho}
\eta^{\nu \kappa} \eta^{\sigma \rho} (F^{2})^{\lambda \mu } 
&=& -F_{\nu\lambda,\mu} F^{\nu\rho,}_{~~~\rho} (F^{2})^{\lambda \mu } 
,\\
F_{\nu\lambda,\mu} F_{\sigma\kappa,\rho}
\eta^{\mu \sigma} \eta^{\nu \kappa} (F^{2})^{\lambda \rho} 
&=&=-\frac{1}{2}F_{\nu\lambda,\mu}F^{\nu\lambda,\rho} 
(F^{2})^{\mu}_{~\rho} , \\
F_{\nu\lambda,\mu} F_{\sigma\kappa,\rho}
\eta^{\mu \rho} \eta^{\nu \kappa} (F^{2})^{\lambda \sigma} 
&=&-F_{\nu\lambda,\mu}F^{\nu\sigma,\mu}(F^{2})^{\lambda}_{~\sigma},\\
F_{\nu\lambda,\mu} F_{\sigma\kappa,\rho}
\eta^{\mu \kappa} \eta^{\nu \rho} (F^{2})^{\lambda \sigma} 
&=&-F_{\nu\lambda,\mu}F^{\nu\sigma,\mu}(F^{2})^{\lambda}_{~\sigma} 
+\frac{1}{2}F_{\nu\lambda,\mu}F^{\nu\lambda,\rho} 
(F^{2})^{\mu}_{~\rho} , \\
F_{\nu\lambda,\mu} F_{\sigma\kappa,\rho}
\eta^{\nu \sigma} \eta^{\lambda \kappa} (F^{2})^{\mu \rho}
&=&F_{\nu\lambda,\mu}F^{\nu\lambda,\rho} 
(F^{2})^{\mu}_{~\rho},\\
F_{\nu\lambda,\mu} F_{\sigma\kappa,\rho}
\eta^{\nu \mu } \eta^{\kappa \rho} (F^{2})^{\lambda \sigma}
&=&-F_{\nu\mu,}^{~~~\mu} F_{\sigma\rho,}^{~~~\rho}
(F^{2})^{\nu \sigma}.
\end{eqnarray}
After expanding $tr\langle x|U(\tau)|x\rangle_{0}$ in 
Eq.~(\ref{tau5fer}) in powers of $\tau$ up to the terms of 
order $\tau^3$ and substituting the obtained expression in the 
definition of the effective action, we arrive at the following 
two-derivative correction of the order $1/m^6$,
\begin{eqnarray} 
&&{\cal L}^{(3+1)spin}_{1/m^6}=-\frac{11\alpha^2}{630 m^6} 
F^{\beta \gamma}F_{\beta \gamma}
F^{\nu \lambda}F_{\nu\lambda,\mu}^{~~~~\mu}
+\frac{4\alpha^2}{315 m^6} (F^{2})^{\mu \kappa}
F^{\nu \lambda} F_{\nu\lambda,\mu\kappa}\nonumber\\
&&+\frac{\alpha^2}{270 m^6} F^{\beta \gamma}F_{\beta \gamma}
\left( \frac{7}{2}F^{\nu \lambda,\mu}F_{\nu\lambda,\mu}-
F^{\nu \lambda,}_{~~~\lambda}F_{\nu\mu,}^{~~~\mu}
\right)
\nonumber\\
&&-\frac{2\alpha^2}{945 m^6} \Bigg(
2 F_{\nu\lambda,\mu} F^{\nu\rho,}_{~~~\rho} (F^{2})^{\lambda \mu } 
-2 F_{\nu\lambda,\mu}F^{\nu\lambda,\rho} (F^{2})^{\mu}_{~\rho} 
-37 F_{\nu\lambda,\mu}F^{\nu\sigma,\mu}(F^{2})^{\lambda}_{~\sigma}
+ F_{\nu\mu,}^{~~~\mu} F_{\sigma\rho,}^{~~~\rho} (F^{2})^{\nu \sigma}
\Bigg)\nonumber\\
&&+\frac{\alpha^2}{630 m^6} 
 F_{\nu\lambda,\mu} F_{\sigma\kappa,\rho} \Bigg(
 38 \eta^{\mu \rho} F^{\lambda \sigma} F^{\nu \kappa} 
-12 \eta^{\nu \kappa} F^{\lambda \rho} F^{\mu \sigma} 
+47 \eta^{\mu \rho} F^{\nu \lambda} F^{\sigma \kappa} 
+16 \eta^{\kappa \rho} F^{\lambda \sigma} F^{\nu \mu } 
\Bigg) ,   
\end{eqnarray}
to the one-loop effective action in spinor QED in $3+1$ dimensions, 
and the correction of the order $1/m^7$,
\begin{eqnarray} 
&&{\cal L}^{(2+1)spin}_{1/m^7}=-\frac{11\alpha^2\pi}{336 m^7} 
F^{\beta \gamma}F_{\beta \gamma}
F^{\nu \lambda}F_{\nu\lambda,\mu}^{~~~~\mu}
+\frac{\alpha^2\pi}{42 m^7}(F^{2})^{\mu \kappa} 
F^{\nu \lambda} F_{\nu\lambda,\mu\kappa}\nonumber\\
&&+\frac{\alpha^2\pi}{144 m^7} F^{\beta \gamma}F_{\beta \gamma}
\left( \frac{7}{2}F^{\nu \lambda,\mu}F_{\nu\lambda,\mu}-
F^{\nu \lambda,}_{~~~\lambda}F_{\nu\mu,}^{~~~\mu}
\right) \nonumber\\
&&-\frac{\alpha^2\pi}{252 m^7} \Bigg(
2 F_{\nu\lambda,\mu} F^{\nu\rho,}_{~~~\rho} (F^{2})^{\lambda \mu } 
-2 F_{\nu\lambda,\mu}F^{\nu\lambda,\rho} (F^{2})^{\mu}_{~\rho} 
-37 F_{\nu\lambda,\mu}F^{\nu\sigma,\mu}(F^{2})^{\lambda}_{~\sigma}
+ F_{\nu\mu,}^{~~~\mu} F_{\sigma\rho,}^{~~~\rho} (F^{2})^{\nu \sigma}
\Bigg)\nonumber\\
&&+\frac{\alpha^2\pi}{336 m^7} 
 F_{\nu\lambda,\mu} F_{\sigma\kappa,\rho} \Bigg(
 38 \eta^{\mu \rho} F^{\lambda \sigma} F^{\nu \kappa} 
-12 \eta^{\nu \kappa} F^{\lambda \rho} F^{\mu \sigma} 
+47 \eta^{\mu \rho} F^{\nu \lambda} F^{\sigma \kappa} 
+16 \eta^{\kappa \rho} F^{\lambda \sigma} F^{\nu \mu } 
\Bigg),    
\end{eqnarray}
to the effective action in $2+1$ dimensions.
It turns out that these latter can be further simplified.
Indeed, after integrating by parts, the results can be 
expressed through the following seven Lorentz scalars,
\begin{eqnarray} 
L_{1}&=&F^{\beta \gamma}F_{\beta \gamma}
F^{\nu \lambda}F_{\nu\lambda,\mu}^{~~~~\mu} \\
L_{2}&=&F^{\beta \gamma}F_{\beta \gamma}
F^{\nu \lambda,\mu}F_{\nu\lambda,\mu} \\
L_{3}&=&F^{\beta \gamma}F_{\beta \gamma}
F^{\nu \lambda,}_{~~~\lambda}F_{\nu\mu,}^{~~~\mu} \\
L_{4}&=&F_{\nu\lambda,\mu\kappa} F^{\nu \lambda} 
(F^{2})^{\mu \kappa} \\
L_{5}&=& F^{\kappa \nu} F_{\nu\lambda,\mu} F^{\lambda \sigma} 
F_{\sigma\kappa,}^{~~~\mu} \\
L_{6}&=&F_{\nu\lambda,\mu}^{~~~~\mu} (F^{3})^{\nu\lambda} \\
L_{7}&=& F_{\nu \mu,}^{~~~\mu} F_{\lambda \rho,}^{~~~\rho}
(F^{2})^{\nu \lambda}.
\end{eqnarray}
Thus, the final results in $3+1$ and in $2+1$ dimensions 
read
\begin{eqnarray} 
{\cal L}^{(3+1)spin}_{1/m^6}&=&
  -\frac{16\alpha^2}{315m^6} L_{1}
  -\frac{8\alpha^2}{315m^6} L_{2}
  +\frac{2\alpha^2}{315m^6} L_{3}
  -\frac{\alpha^2}{945m^6} L_{4}\nonumber\\
&&-\frac{11\alpha^2}{945m^6} L_{5}
  -\frac{26\alpha^2}{945m^6} L_{6}
  +\frac{4\alpha^2}{189m^6} L_{7},\\
{\cal L}^{(2+1)spin}_{1/m^7}&=&
  -\frac{2\alpha^2\pi}{21m^6} L_{1}
  -\frac{\alpha^2\pi}{21m^6} L_{2}
  +\frac{\alpha^2\pi}{84m^6} L_{3}
  -\frac{\alpha^2\pi}{504m^6} L_{4}\nonumber\\
&&-\frac{11\alpha^2\pi}{504m^6} L_{5}
  -\frac{13\alpha^2\pi}{252m^6} L_{6}
  +\frac{5\alpha^2\pi}{126m^6} L_{7},
\end{eqnarray}
respectively. This should be compared with the result of
\cite{Dicus} (see Eq.~(14) there). Notice that the photon field
in Ref.~\cite{Dicus} desribes on-shell quanta, and, as a
result, the terms containing $L_{1}$, $L_{3}$, $L_{6}$ and
$L_{7}$ do not appear (they are proportional to $k^2=0$).

In a similar way, in the case of scalar QED we obtain the
following expression for the expansion of Eq.~(\ref{eq:trUbos}), 
\begin{eqnarray} 
&&\langle x|U_{bos}(\tau)|x\rangle\simeq
\langle x|U_{bos}(\tau)|x\rangle_{0} 
\Bigg[1 -\frac{ie^2\tau^3}{30} 
F^{\nu \lambda}F_{\nu\lambda,\mu}^{~~~~\mu}\nonumber\\
&&-\frac{ie^2\tau^3}{180}\left( 
4F^{\nu \lambda,\mu}F_{\nu\lambda,\mu}+
F^{\nu \lambda,}_{~~~\lambda}F_{\nu\mu,}^{~~~\mu}
\right)+\frac{ie^4\tau^5}{840} F_{\nu\lambda,\mu\kappa}
\left(4 {\cal F} \eta^{\mu \kappa} F^{\nu \lambda} 
  + 2 F^{\nu \lambda} (F^{2})^{\mu \kappa}
\right)\nonumber\\
&+&\frac{ie^4\tau^5}{7560} \Bigg(
8 F_{\nu\lambda,\mu} F^{\nu\rho,}_{~~~\rho} (F^{2})^{\lambda \mu } 
+13 F_{\nu\lambda,\mu}F^{\nu\lambda,\rho} (F^{2})^{\mu}_{~\rho} 
+20 F_{\nu\lambda,\mu}F^{\nu\sigma,\mu}(F^{2})^{\lambda}_{~\sigma}
+ 4 F_{\nu\mu,}^{~~~\mu} F_{\sigma\rho,}^{~~~\rho} (F^{2})^{\nu \sigma}
\Bigg)\nonumber\\
&+&\frac{ie^4\tau^5}{5040}
 F_{\nu\lambda,\mu} F_{\sigma\kappa,\rho} \Bigg(
 8 \eta^{\mu \rho} F^{\lambda \sigma} F^{\nu \kappa} 
-4 \eta^{\nu \kappa} F^{\lambda \rho} F^{\mu \sigma} 
-17 \eta^{\mu \rho} F^{\nu \lambda} F^{\sigma \kappa} 
+24 \eta^{\kappa \rho} F^{\lambda \sigma} F^{\nu \mu } 
\Bigg)\Bigg],
\label{tau5bos}
\end{eqnarray}
leading to the $1/m^6$ correction to the effective 
Lagrangian density,
\begin{eqnarray} 
&&{\cal L}^{(3+1)scal}_{1/m^6}=-\frac{\alpha^2}{126 m^6} 
F^{\beta \gamma}F_{\beta \gamma}
F^{\nu \lambda}F_{\nu\lambda,\mu}^{~~~~\mu}
+\frac{\alpha^2}{210 m^6} (F^{2})^{\mu \kappa} 
F^{\nu \lambda} F_{\nu\lambda,\mu\kappa}\nonumber\\
&&-\frac{\alpha^2}{1080 m^6} F^{\beta \gamma}F_{\beta \gamma}
\left( 4F^{\nu \lambda,\mu}F_{\nu\lambda,\mu}+
F^{\nu \lambda,}_{~~~\lambda}F_{\nu\mu,}^{~~~\mu}
\right)\nonumber\\
&&+\frac{\alpha^2}{3780 m^6} \Bigg(
8 F_{\nu\lambda,\mu} F^{\nu\rho,}_{~~~\rho} (F^{2})^{\lambda \mu } 
+13 F_{\nu\lambda,\mu}F^{\nu\lambda,\rho} (F^{2})^{\mu}_{~\rho} 
+20 F_{\nu\lambda,\mu}F^{\nu\sigma,\mu}(F^{2})^{\lambda}_{~\sigma}
+ 4 F_{\nu\mu,}^{~~~\mu} F_{\sigma\rho,}^{~~~\rho} (F^{2})^{\nu \sigma}
\Bigg)\nonumber\\
&&+\frac{\alpha^2}{2520 m^6} 
 F_{\nu\lambda,\mu} F_{\sigma\kappa,\rho} \Bigg(
 8 \eta^{\mu \rho} F^{\lambda \sigma} F^{\nu \kappa} 
-4 \eta^{\nu \kappa} F^{\lambda \rho} F^{\mu \sigma} 
-17 \eta^{\mu \rho} F^{\nu \lambda} F^{\sigma \kappa} 
+24 \eta^{\kappa \rho} F^{\lambda \sigma} F^{\nu \mu } 
\Bigg) ,   
\end{eqnarray}
in $3+1$ dimensions, and the $1/m^7$ correction, 
\begin{eqnarray} 
&&{\cal L}^{(2+1)scal}_{1/m^7}=\frac{5\alpha^2\pi}{336 m^7} 
F^{\beta \gamma}F_{\beta \gamma}
F^{\nu \lambda}F_{\nu\lambda,\mu}^{~~~~\mu}
-\frac{\alpha^2\pi}{112 m^7} (F^{2})^{\mu \kappa} 
F^{\nu \lambda} F_{\nu\lambda,\mu\kappa}\nonumber\\
&&+\frac{\alpha^2\pi}{576 m^7} F^{\beta \gamma}F_{\beta \gamma}
\left( 4F^{\nu \lambda,\mu}F_{\nu\lambda,\mu}+
F^{\nu \lambda,}_{~~~\lambda}F_{\nu\mu,}^{~~~\mu}
\right)\nonumber\\
&&-\frac{\alpha^2\pi}{2016 m^7} \Bigg(
8 F_{\nu\lambda,\mu} F^{\nu\rho,}_{~~~\rho} (F^{2})^{\lambda \mu } 
+13 F_{\nu\lambda,\mu}F^{\nu\lambda,\rho} (F^{2})^{\mu}_{~\rho} 
+20 F_{\nu\lambda,\mu}F^{\nu\sigma,\mu}(F^{2})^{\lambda}_{~\sigma}
+ 4 F_{\nu\mu,}^{~~~\mu} F_{\sigma\rho,}^{~~~\rho} (F^{2})^{\nu \sigma}
\Bigg)\nonumber\\
&&-\frac{\alpha^2\pi}{1344 m^7} 
 F_{\nu\lambda,\mu} F_{\sigma\kappa,\rho} \Bigg(
 8 \eta^{\mu \rho} F^{\lambda \sigma} F^{\nu \kappa} 
-4 \eta^{\nu \kappa} F^{\lambda \rho} F^{\mu \sigma} 
-17 \eta^{\mu \rho} F^{\nu \lambda} F^{\sigma \kappa} 
+24 \eta^{\kappa \rho} F^{\lambda \sigma} F^{\nu \mu } 
\Bigg) ,   
\end{eqnarray}
in $2+1$ dimensions.

Up to a divergence, the derived corrections to the effective action
are equivalent to 
\begin{eqnarray} 
{\cal L}^{(3+1)scal}_{1/m^6}&=&
  -\frac{13\alpha^2}{2520m^6} L_{1}
  -\frac{\alpha^2}{840m^6} L_{2}
  -\frac{\alpha^2}{2520m^6} L_{3}
  +\frac{\alpha^2}{1890m^6} L_{4}\nonumber\\
&&+\frac{\alpha^2}{3780m^6} L_{5}
  -\frac{11\alpha^2}{3780m^6} L_{6}
  +\frac{\alpha^2}{1890m^6}L_{7},\\ 
{\cal L}^{(2+1)scal}_{1/m^7}&=&
   \frac{13\alpha^2}{1344m^6} L_{1}
  +\frac{\alpha^2}{448m^6} L_{2}
  +\frac{\alpha^2}{1344m^6} L_{3}
  -\frac{\alpha^2}{1008m^6} L_{4}\nonumber\\
&&-\frac{\alpha^2}{2016m^6} L_{5}
  +\frac{11\alpha^2}{2016m^6} L_{6}
  -\frac{\alpha^2}{1008m^6} L_{7},
\end{eqnarray} in $3+1$ and $2+1$ dimensions, respectively. Here
we used the same  seven scalars as in the case of spinor QED above.

\section{Coefficient functions which appear in 
purely electric and purely magnetic cases}
\label{appC}

In this appendix we list the formulas, similar to that in 
Eq.~(\ref{eq:exmpl}), which appear in the course of reduction
the general expression for the derivative contribution
to the case of a pure magnetic (electric) field background. 
These are
\begin{eqnarray} 
&&F_{\nu \lambda,\mu \kappa} \sum_{j,l} C^{W}(f_{j},f_{l})
A^{ \lambda \nu }_{(j)}A^{\mu \kappa }_{(l)}
=-2iC^{W}(\bar{\alpha},\bar{\alpha}) 
\sum\limits_{i=1}^{2}\partial_{i} \partial_{i} B , \\
&&F_{\nu \lambda,\mu \kappa} \sum_{j,l} C^{V}(f_{j},f_{l})
\left(A^{\nu \lambda}_{(j)}A^{\mu \kappa }_{(l)}+
2 A^{\nu \mu}_{(j)}A^{\lambda \kappa }_{(l)}
\right)=4iC^{V}(\bar{\alpha},\bar{\alpha}) 
\sum\limits_{i=1}^{2}\partial_{i} \partial_{i} B , \\
&&F_{\nu \lambda,\mu}F_{\sigma \kappa,\rho} \sum_{j,l,k} 
C_{1}^{WW}(f_{j},f_{l},f_{k})
A^{\kappa \sigma}_{(j)}A^{\lambda \nu}_{(l)}A^{\mu \rho}_{(k)}
=4C_{1}^{WW}(\bar{\alpha},\bar{\alpha},\bar{\alpha}) 
\sum\limits_{i=1}^{2}\left(\partial_{i}B\right)^2 , \\
&&F_{\nu \lambda,\mu}F_{\sigma \kappa,\rho} \sum_{j,l,k} 
C_{2}^{WW}(f_{j},f_{l},f_{k})
A^{\kappa \lambda}_{(j)}A^{\sigma \nu}_{(l)}A^{\mu \rho}_{(k)}
=-2C_{2}^{WW}(\bar{\alpha},-\bar{\alpha},\bar{\alpha}) 
\sum\limits_{i=1}^{2}\left(\partial_{i}B\right)^2 , \\
&&F_{\nu \lambda,\mu}F_{\sigma \kappa,\rho} \sum_{j,l,k} 
C_{1}^{VW}(f_{j},f_{l},f_{k})
A^{\sigma \kappa}_{(j)}\left(
A^{\nu \lambda}_{(l)}A^{\mu \rho}_{(k)}
+A^{\nu \mu}_{(l)}A^{\lambda \rho}_{(k)}
\right)\nonumber\\
&&=2\left( C_{1}^{VW}(\bar{\alpha},\bar{\alpha},\bar{\alpha}) 
+2C_{1}^{VW}(\bar{\alpha},\bar{\alpha},-\bar{\alpha}) 
\right)\sum\limits_{i=1}^{2}\left(\partial_{i}B\right)^2 , \\
&&F_{\nu \lambda,\mu}F_{\sigma \kappa,\rho} \sum_{j,l,k} 
C_{2}^{VW}(f_{j},f_{l},f_{k})
A^{\sigma \kappa}_{(j)}A^{\nu \rho}_{(l)}A^{\lambda \mu}_{(k)}
=2C_{2}^{VW}(\bar{\alpha},\bar{\alpha},\bar{\alpha}) 
\sum\limits_{i=1}^{2}\left(\partial_{i}B\right)^2 , \\
&&F_{\nu \lambda,\mu}F_{\sigma \kappa,\rho} \sum_{j,l,k} 
C_{1}^{VV}(f_{j},f_{l},f_{k})\left(
A^{\nu \lambda}_{(j)}A^{\kappa \sigma}_{(l)}A^{\mu \rho}_{(k)}
+A^{\nu \mu}_{(j)}A^{\kappa\rho}_{(l)}A^{\lambda \sigma}_{(k)}
+2A^{\nu \lambda}_{(j)}A^{\kappa \rho}_{(l)}A^{\mu \sigma}_{(k)}
\right)\nonumber\\
&&=\left(
4C_{1}^{VV}(\bar{\alpha},\bar{\alpha},\bar{\alpha}) 
-4C_{1}^{VV}(-\bar{\alpha},\bar{\alpha},\bar{\alpha}) 
-C_{1}^{VV}(\bar{\alpha},-\bar{\alpha},\bar{\alpha}) 
\right)\sum\limits_{i=1}^{2}\left(\partial_{i}B\right)^2 , \\
&&F_{\nu \lambda,\mu}F_{\sigma \kappa,\rho} \sum_{j,l,k} 
C_{2}^{VV}(f_{j},f_{l},f_{k})\left(
A^{\nu \sigma}_{(j)}A^{\kappa \lambda}_{(l)}A^{\mu \rho}_{(k)}
+A^{\nu \rho}_{(j)}A^{\kappa\mu}_{(l)}A^{\lambda \sigma}_{(k)}
+2A^{\nu \sigma}_{(j)}A^{\kappa \mu}_{(l)}A^{\lambda\rho}_{(k)}
\right)\nonumber\\ &&=-\left(
4C_{2}^{VV}(\bar{\alpha},\bar{\alpha},\bar{\alpha}) 
-C_{2}^{VV}(-\bar{\alpha},\bar{\alpha},\bar{\alpha}) 
\right)\sum\limits_{i=1}^{2}\left(\partial_{i}B\right)^2 , \\
&&F_{\nu \lambda,\mu}F_{\sigma \kappa,\rho} \sum_{j,l,k} 
C_{3}^{VV}(f_{j},f_{l},f_{k})\left(
A^{\nu \lambda}_{(j)}A^{\kappa \mu}_{(l)}A^{\sigma \rho}_{(k)}
+A^{\kappa \rho}_{(j)}A^{\nu\sigma}_{(l)}A^{\lambda \mu}_{(k)}
\right)\nonumber\\ &&=-\left(
4C_{3}^{VV}(\bar{\alpha},\bar{\alpha},\bar{\alpha}) 
-C_{3}^{VV}(\bar{\alpha},-\bar{\alpha},\bar{\alpha}) 
\right)\sum\limits_{i=1}^{2}\left(\partial_{i}B\right)^2 , \\
&&C_{4}^{VV}(f_{j},f_{l},f_{k})
A^{\nu \kappa}_{(j)}A^{\lambda \mu}_{(l)}A^{\sigma \rho}_{(k)}
=C_{4}^{VV}(\bar{\alpha},\bar{\alpha},\bar{\alpha}) 
\sum\limits_{i=1}^{2}\left(\partial_{i}B\right)^2 , \\
&&F_{\nu \lambda,\mu}F_{\sigma \kappa,\rho} \sum_{j,l,k} 
C_{5}^{VV}(f_{j},f_{l},f_{k})
A^{\nu \kappa}_{(j)} \left( A^{\lambda \sigma}_{(l)}A^{\mu \rho}_{(k)}
+A^{\lambda \rho}_{(l)}A^{\mu \sigma}_{(k)}
\right)\nonumber\\  &&=\left(
C_{5}^{VV}(\bar{\alpha},-\bar{\alpha},\bar{\alpha}) 
+2C_{5}^{VV}(-\bar{\alpha},\bar{\alpha},\bar{\alpha}) 
\right)\sum\limits_{i=1}^{2}\left(\partial_{i}B\right)^2 , \\
\end{eqnarray} 
where $\bar{\alpha}=\sqrt{2{\cal F}}$.

The latter expressions contain the coefficient functions 
from Appendix~\ref{appA} calculated for a particular values 
of their arguments. The convenient representation for them
reads
\begin{eqnarray} 
C^{V}(\bar{\alpha},\bar{\alpha}) &=& \frac{\tau^{2}}{2\omega } 
\left( 3\omega^{2}H^{2} + 3H -1 \right), \\ 
C^{W}(\bar{\alpha},\bar{\alpha}) &=& \tau^{2} \tanh(\omega) H,\\ 
C^{WW}_{1}(\bar{\alpha},\bar{\alpha},\bar{\alpha}) &=& \frac{\tau 
^{3}}{8} \tanh^{2}(\omega ) H, \\ 
C^{WW}_{2}(\bar{\alpha},-\bar{\alpha},\bar{\alpha}) &=& - 
\frac{\tau ^{3}}{4} \left( 1 - \tanh^{2}(\omega ) \right) H, \\ 
C^{VW}_{1}(\bar{\alpha},\bar{\alpha},\bar{\alpha}) &=& - 
\frac{\tau ^{3}}{4} \tanh(\omega ) \omega H^{2} ,\\ 
C^{VW}_{1}(\bar{\alpha},\bar{\alpha},-\bar{\alpha}) &=& - 
\frac{\tau ^{3}}{4\omega } \tanh(\omega ) \left( 2\omega ^{2}H^{2}+ 
3H -1 \right) ,\\ 
C^{VW}_{2}(\bar{\alpha},\bar{\alpha},\bar{\alpha}) &=& - 
\frac{\tau ^{3}}{4\omega } \tanh(\omega ) \left( \omega ^{2}H^{2}+ 
3H -1 \right) ,\\ 
C^{VV}_{1}(\bar{\alpha},\bar{\alpha},\bar{\alpha}) &=& \frac{\tau 
^{3}}{4\omega^{2} } \left( 4\omega ^{4}H^{3}+ 7\omega ^{2}H^{2} 
-2\omega ^{2}H +3H -1 \right) ,\\ 
C^{VV}_{1}(-\bar{\alpha},\bar{\alpha},\bar{\alpha}) &=& 
-\frac{\tau ^{3}}{2\omega^{2} } \left( 4\omega ^{4}H^{3}+10\omega 
^{2}H^{2} -3\omega ^{2}H +6H -2 \right) ,\\ 
C^{VV}_{1}(\bar{\alpha},-\bar{\alpha},\bar{\alpha}) &=& 
-\frac{\tau ^{3}}{4\omega^{2} } \left( 2\omega ^{4}H^{3}- \omega 
^{2}H^{2}-3H +1 \right) ,\\ 
C^{VV}_{2}(\bar{\alpha},\bar{\alpha},\bar{\alpha}) &=& \frac{\tau 
^{3}}{2\omega^{2} } \left( 2\omega ^{4}H^{3}+5\omega ^{2}H^{2} 
-2\omega ^{2}H +3H -1 \right) ,\\ 
C^{VV}_{2}(-\bar{\alpha},\bar{\alpha},\bar{\alpha}) &=& 
-\frac{\tau ^{3}}{4\omega^{2} } \left( 2\omega ^{4}H^{3}+11\omega 
^{2}H^{2} -2\omega ^{2}H +9H -3 \right) ,\\ 
C^{VV}_{3}(\bar{\alpha},\bar{\alpha},\bar{\alpha}) &=& \frac{\tau 
^{3}}{4\omega^{2} } \left( 4\omega ^{4}H^{3}+13\omega ^{2}H^{2} 
-4\omega ^{2}H +9H -3 \right) ,\\ 
C^{VV}_{3}(\bar{\alpha},-\bar{\alpha},\bar{\alpha}) &=& 
-\frac{\tau ^{3}}{2\omega^{2} } \left( \omega ^{4}H^{3}+ 4\omega 
^{2}H^{2} - \omega ^{2}H +3H -1 \right) ,\\ 
C^{VV}_{4}(\bar{\alpha},\bar{\alpha},\bar{\alpha}) &=& -\frac{\tau 
^{3}}{4\omega^{2} } \left( 2\omega ^{4}H^{3}+ 5\omega ^{2}H^{2} 
-2\omega ^{2}H +3H -1 \right) ,\\ 
C^{VV}_{5}(\bar{\alpha},-\bar{\alpha},\bar{\alpha}) &=& 
-\frac{\tau ^{3}}{4\omega^{2} } \left( 2\omega ^{4}H^{3}- \omega 
^{2}H^{2} -2\omega ^{2}H -3H +1 \right) ,\\ 
C^{VV}_{5}(-\bar{\alpha},\bar{\alpha},\bar{\alpha}) &=& 
-\frac{\tau ^{3}}{\omega^{2} } \left( 2\omega ^{4}H^{3}+5\omega 
^{2}H^{2} -2\omega ^{2}H +3H -1 \right), 
\end{eqnarray} 
where, by definition, $\omega=e\bar{\alpha}\tau$ and $H=H(\omega)$.

\section{Special function representation for the integrals which
appear in the purely electric and purely magnetic cases}
\label{appD}

In the main text, we saw that the calculation of the effective 
action for spinor QED in an external magnetic field reduces to 
evaluating the following integral (with $\mu=1/2$ in $2+1$ 
dimensions, and $\mu=0$ in $3+1$ dimensions)
\begin{eqnarray} 
I^{(spin)}(\sigma;\mu)&=& 
\int\limits_{0}^{\infty} d\omega\omega^{\mu-1}
e^{-\sigma\omega}\frac{d^3}{d\omega^3}
\left( \omega \coth\omega \right)= 
\int\limits_{0}^{\infty} d\omega\omega^{\mu-1}     
e^{-\sigma\omega}\frac{d^3}{d\omega^3}
\left( \omega \coth\omega -1-\frac{\omega^2}{3}\right) 
\nonumber \\
&=& -\int\limits_{0}^{\infty} d\omega 
\left( \omega \coth\omega -1-\frac{\omega^2}{3}\right)
\frac{d^3}{d\omega^3}
\left( \omega^{\mu-1} e^{-\sigma\omega}\right)
\nonumber\\
&=&\int\limits_{0}^{\infty}d\omega\omega^{\mu-1} e^{-\sigma\omega}
\left(\coth\omega -\frac{1}{\omega}-\frac{\omega}{3}\right)
\left(\frac{(3-\mu)(2-\mu)(1-\mu)}{\omega^2} \right.\nonumber \\
&+&\left.\frac{3\sigma(2-\mu)(1-\mu)}{\omega}+3\sigma^2(1-\mu)
+\sigma^3\omega\right),
\label{eq:d3} 
\end{eqnarray} 
where we integrated by parts (to avoid divergences as $\omega\to 
0$ we subtracted the first two terms of the hyperbolic cotangent 
asymptotes). For large enough values of the parameter $\mu$, one 
can apply the following table integrals \cite{Ryzh}
\begin{eqnarray} 
\int\limits_{0}^{\infty} d\omega\omega^{\mu-1}
e^{-\sigma\omega}\coth\omega &=& \Gamma(\mu) 
\left[2^{1-\mu}\zeta\left(\mu,1+\frac{\sigma}{2}\right)
+\sigma^{-\mu}\right], \label{eq:d4} \\
\int\limits_{0}^{\infty} d\omega\omega^{\mu-1} e^{-\sigma\omega} 
&=& \sigma^{-\mu} \Gamma(\mu).\label{eq:d5} 
\end{eqnarray} 
Thus, the integral in Eq.~(\ref{eq:d3}) yields
\begin{eqnarray} 
I^{(spin)}(\sigma;\mu)&=& 2^{-\mu} \Gamma(\mu+1)
\left[\sigma^3\zeta\left(\mu+1,1+\frac{\sigma}{2}\right) 
+6\sigma^2\frac{1-\mu}{\mu}
\zeta\left(\mu,1+\frac{\sigma}{2}\right) \right.\nonumber\\
&-& \left. 12\sigma\frac{2-\mu}{\mu}
\zeta\left(\mu-1,1+\frac{\sigma}{2}\right) 
+8\frac{3-\mu}{\mu}
\zeta\left(\mu-2,1+\frac{\sigma}{2}\right)\right].
\label{eq:d6}
\end{eqnarray} 
As one can easily check, the original integral in Eq.~(\ref{eq:d3})
is well defined for $\mu>-1$. Therefore, the last expression should
allow a well defined analytical continuation to the whole that range 
of values of $\mu$. Notice that this should be true even despite the 
fact that the intermediate integrals, as in Eqs.~(\ref{eq:d4}) and 
(\ref{eq:d5}), may not be well defined for all values $\mu>-1$.
In particular, by an analytical continuation, we obtain the results 
for the values of $\mu$ which are of interest,
\begin{eqnarray} 
I^{(spin)}\left(\sigma;\frac{1}{2}\right)&=& 2\sqrt{2\pi} 
\left[
5\zeta\left(-\frac{3}{2},1+\frac{\sigma}{2}\right)
-9\frac{\sigma}{2}
\zeta\left(-\frac{1}{2},1+\frac{\sigma}{2}\right)
\right.\nonumber\\
&+&\left. 3\left(\frac{\sigma}{2}\right)^2
\zeta\left(\frac{1}{2},1+\frac{\sigma}{2}\right) 
+\left(\frac{\sigma}{2}\right)^3
\zeta\left(\frac{3}{2},1+\frac{\sigma}{2}\right)\right], 
\label{eq:d7}\\
I^{(spin)}\left(\sigma;0\right)&=& \frac{11}{6}\sigma^3
+\sigma^2-\frac{1}{3}\sigma-\sigma^3
\psi\left(1+\frac{\sigma}{2}\right)
+6\sigma^2\left[\ln\Gamma\left(1+\frac{\sigma}{2}\right)
-\ln\sqrt{2\pi}\right] \nonumber\\
&-& 24\sigma\zeta'\left(-1,1+\frac{\sigma}{2}\right)
+24\zeta'\left(-2,1+\frac{\sigma}{2}\right),
\label{eq:d8}
\end{eqnarray} 
where the prime denotes the derivative of zeta function with 
respect to its first argument. In derivation of the second 
expression we used the following identities \cite{Ryzh}
\begin{eqnarray} 
\zeta(-1,q)=-\frac{q^2}{2}+\frac{q}{2}-\frac{1}{12},
&&\qquad \zeta(0,q)=\frac{1}{2}-q, \nonumber\\
\zeta(-2,q)=-\frac{q^3}{3}+\frac{q^2}{2}-\frac{q}{6}, 
&&\qquad \zeta'(0,q)\equiv\left.\frac{\partial\zeta(z,q)}
{\partial z}\right|_{z=0}=\ln\Gamma(q)-\ln\sqrt{2\pi},\nonumber\\
\lim_{z\to 1}\left(\zeta(z,q)-\frac{1}{z-1}\right)=-\psi(q).&&
\end{eqnarray} 

In the case of scalar QED, we come to the integral (again, with 
$\mu=1/2$ in $2+1$ dimensions, and $\mu=0$ in $3+1$ dimensions)
\begin{eqnarray} 
&&I^{(scal)}(\sigma;\mu)= 
\int\limits_{0}^{\infty} 
d\omega\omega^{\mu-1}e^{-\sigma\omega}
\left(\frac{d^3}{d\omega^3}+\frac{d}{d\omega}\right)
\frac{\omega}{\sinh\omega } \nonumber\\
&=& \int\limits_{0}^{\infty} 
d\omega\omega^{\mu-1} e^{-\sigma\omega}\left[
\frac{d^3}{d\omega^3} \left(\frac{\omega}{\sinh\omega } 
-1+\frac{\omega^2}{6}\right)+\frac{d}{d\omega}
\left( \frac{\omega}{\sinh\omega } -1\right)\right]
 \nonumber\\ &=& 
-\int\limits_{0}^{\infty} d\omega \left[ 
\left(\frac{\omega}{\sinh\omega } 
-1+\frac{\omega^2}{6}\right)
\frac{d^3}{d\omega^3}
+\left(\frac{\omega}{\sinh\omega } -1\right)
\frac{d}{d\omega}\right]
\left(\omega^{\mu-1} e^{-\sigma\omega}\right)
 \nonumber\\ 
&=&\int\limits_{0}^{\infty}d\omega\omega^{\mu-1} 
e^{-\sigma\omega}\left[ 
\left(\frac{1}{\sinh\omega } 
-\frac{1}{\omega}+\frac{\omega}{6}\right)
\left(\frac{(3-\mu)(2-\mu)(1-\mu)}{\omega^2} 
\right.\right.\nonumber\\ &+&\left.\left. 
\frac{3\sigma(2-\mu)(1-\mu)}{\omega} 
+3\sigma^2(1-\mu)+\sigma^3\omega\right) 
+\left(\frac{1}{\sinh\omega } 
-\frac{1}{\omega}\right) \left(1-\mu 
+\sigma\omega\right) \right],
\label{eq:d11} 
\end{eqnarray} 
where we integrated by parts as in the spinor case. In addition to 
the table integral in (\ref{eq:d5}), we need also the following 
one,
\begin{eqnarray} 
\int\limits_{0}^{\infty} \frac{d\omega\omega^{\mu-1} 
e^{-\sigma\omega}}{\sinh\omega }&=& 2^{1-\mu} \Gamma(\mu) 
\zeta\left(\mu,\frac{1+\sigma}{2}\right). 
\label{eq:d12} 
\end{eqnarray} 
Thus, we obtain
\begin{eqnarray} 
I^{(scal)}(\sigma;\mu)&=& 2^{-\mu} \Gamma(\mu+1)
\left[\sigma(1+\sigma^2)\zeta\left(\mu+1,\frac{1+\sigma}{2}\right) 
+2(1+3\sigma^2)\frac{1-\mu}{\mu}
\zeta\left(\mu,\frac{1+\sigma}{2}\right) \right. \nonumber\\ 
&-&\left.12\sigma\frac{2-\mu}{\mu}
\zeta\left(\mu-1,\frac{1+\sigma}{2}\right) 
+8\frac{3-\mu}{\mu}
\zeta\left(\mu-2,\frac{1+\sigma}{2}\right)\right].
\label{eq:d13}
\end{eqnarray} 
And, finally, by analytical continuation, we obtain the 
results for two values of $\mu$ that are of interest, 
\begin{eqnarray} 
I^{(scal)}\left(\sigma;\frac{1}{2}\right)&=& 
\sqrt{\frac{\pi}{2}} 
\left[
20\zeta\left(-\frac{3}{2},\frac{1+\sigma}{2}\right)
-18\sigma\zeta\left(-\frac{1}{2},\frac{1+\sigma}{2}\right) 
\right. \nonumber\\ &+&\left.(1+3\sigma^2)
\zeta\left(\frac{1}{2},\frac{1+\sigma}{2}\right) 
+\frac{\sigma}{2}(1+\sigma^2)
\zeta\left(\frac{3}{2},\frac{1+\sigma}{2}\right)\right], 
\label{eq:d14}\\
I^{(scal)}\left(\sigma;0\right)&=& \frac{11}{6}\sigma^3
+\frac{7}{6}\sigma-\sigma(1+\sigma^2)
\psi\left(\frac{1+\sigma}{2}\right)
+2(1+3\sigma^2)\left[\ln\Gamma\left(\frac{1+\sigma}{2}\right)
-\ln\sqrt{2\pi}\right] \nonumber\\ 
&-& 24\sigma\zeta'\left(-1,\frac{1+\sigma}{2}\right)
+24\zeta'\left(-2,\frac{1+\sigma}{2}\right).
\label{eq:d15}
\end{eqnarray}

\begin{figure}
\epsfbox{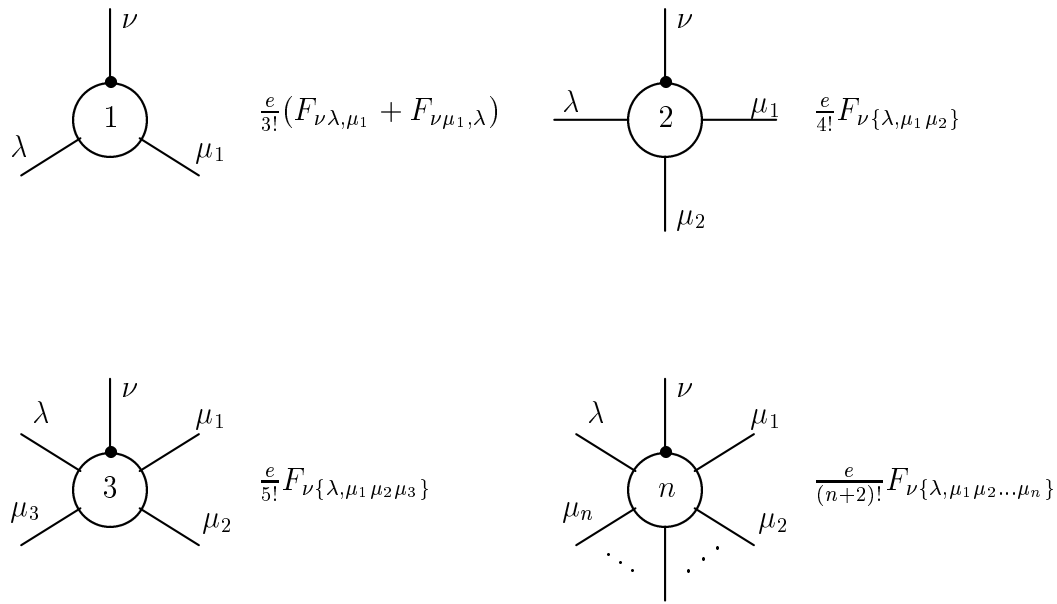}
\label{Figure1}
\caption{Diagrammatic notations for the boson
interaction vertices.  The curly brackets denote symmetrization 
of the type:  $F_{\nu\{\lambda,\mu_1\dots\mu_n\}} 
=F_{\nu\lambda,\mu_1\dots\mu_n} 
+F_{\nu\mu_1,\lambda\dots\mu_n} +\dots 
+F_{\nu\mu_n,\mu_1\dots\lambda}$.}
\end{figure}
\vspace{1in}

\begin{figure}
\epsfbox{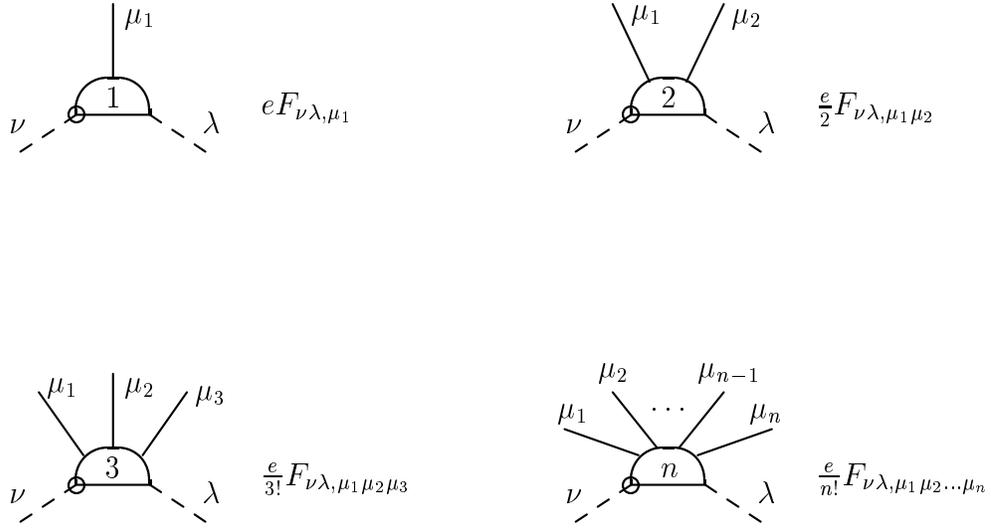}
\label{Figure2}
\caption{Diagrammatic notations for the
fermion-boson  interaction vertices.}
\end{figure}
\vspace{1in}

\begin{figure}
\epsfbox{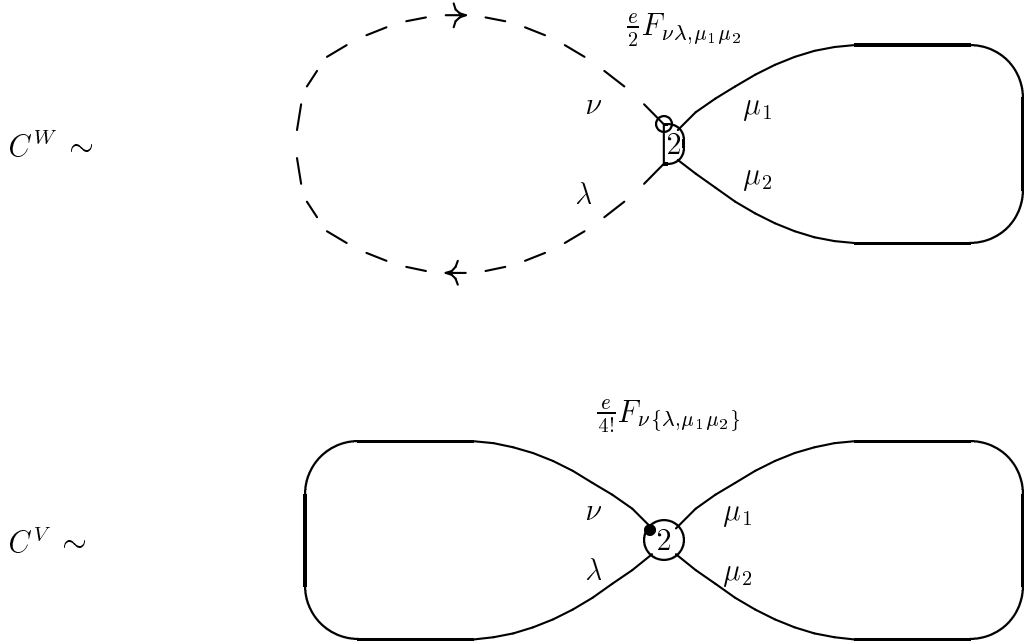}
\label{Figure3}
\caption{Two simplest examples of diagrams related
to the two-derivative terms $C^{W}$ and $C^{V}$ in our general 
expression for spinor QED.}
\end{figure}
\vspace{1in}

\end{document}